\documentclass[12pt]{article}

\usepackage{amsmath,amssymb,epsfig,xspace,epic,array,fullpage,version,amsthm}
\usepackage{setspace}

\newcommand{\Weyl}{\mathcal{W}}
\newcommand{\parb}{\pmb{\partial}}

\def\be{\begin{equation}}
\def\ee{\end{equation}}
\def\bea{\begin{eqnarray}}
\def\eea{\end{eqnarray}}
\def\lb{\label}

\def\H{\mathcal{S}}

\def\gam{\gamma}
\def\d{\delta}
\def\eps{\epsilon}

\def\hsp5{\hspace{5mm}}

\def\Sig{\Sigma}

\def\Om{\Omega}

\def\ptl{\partial}

\def\la{\langle}
\def\ra{\rangle}

\def\be{\begin{equation}}
\def\ee{\end{equation}}
\def\etal{{\em et al.~\/}}

\begin{document} 
\begin{spacing}{1}

\begin{center}
{\large\bf Cosmological models from a dynamical systems
perspective}\\[5mm]
J. Wainwright$^{1}$ and W.C. Lim$^{1,2}$
\\[2mm]
$^{1}$ Department of Applied Mathematics, University of Waterloo,\\
Waterloo, Ontario, Canada N2L 3G1\\
$^{2}$ Department of Mathematics and Statistics, Dalhousie University,\\
Halifax, Nova Scotia, Canada B3H 3J5\\
[2mm]
Email: jwainwri@math.uwaterloo.ca, wclim@mathstat.dal.ca\\[2mm]
\end{center}

\begin{abstract}

It is useful to study the space of all cosmological
models from a dynamical systems perspective, that is, by
formulating the Einstein field equations as a dynamical
system using appropriately normalized variables.  We will
discuss various aspects of this work, the choices of   
normalization factor, multiple representations of models,
the past attractor, nonlinear dynamics in
close-to-Friedmann-Lema\^{\i}tre models, Weyl curvature
dominance, and numerical simulations.
\end{abstract}

\section{Introduction}

The hot big-bang model of modern cosmology is based on the assumption 
that the large scale geometry and dynamics of the universe can be 
described by an exact FLRW model, with more detailed and smaller scale 
physical phenomena, such as density fluctuations or gravitational waves, 
being described by perturbed FL models, i.e.~by solutions of the EFE 
linearized about an exact FL model.

We begin with the premise that it is important to study cosmological 
models more general than perturbed FL models.
In particular one is interested in the set 
$\mathcal{M}_{\rm obs}$, namely, the set of all universes that 
are compatible with current observations (Ellis 2004).
A member of this set must have an epoch, possibly finite, during which the 
model is close to FL;
at early or later times it may deviate significantly from FL.

More generally we are interested in the set $\mathcal{M}$ of \emph{all} 
cosmological models, and in the relationship between $\mathcal{M}$,
$\mathcal{M}_{\rm obs}$ and $\mathcal{M}_{\rm FL}$, the set of 
all FL models.
For example, we want to study the relation between linearly perturbed FL 
models and models that are close to FL in some well-defined, 
gauge-invariant sense, and satisfy the exact EFE.
Are nonlinear effects perhaps significant in close-to-FL models?
Another question that takes one outside the realm of linearly perturbed FL 
models is the detailed nature of the generic cosmological singularity.

With the preceding discussion as motivation 
we consider the class $\mathcal{M}$ of all cosmological models,
whose state at time $t$ can be represented by a vector $\mathbf{X}$ in a 
state space $\mathcal{S}$.
The evolution of a model universe is then described by a curve 
$\mathbf{X}=\mathbf{X}(t)$ in $\mathcal{S}$, called an orbit, that will be 
a solution of a system of first order autonomous evolution equations and
constraints of the form
\begin{gather*}
	\ptl_t \mathbf{X} = \mathbf{F} ( \mathbf{X}, \ptl_i \mathbf{X}, 
		\cdots ),
\\
	 \mathbf{C} ( \mathbf{X}, \ptl_i \mathbf{X}, \cdots )=0,
\end{gather*}
where $\ptl_i$ denotes partial differentiation with respect to the spatial 
coordinates $x^i$, and $\cdots$ denotes possible higher order spatial 
derivatives.
The goal is to choose state variables that remain bounded during the 
evolution of the model and a time variable $t$ such that 
$t\rightarrow-\infty$ at the initial singularity
and $t\rightarrow+\infty$ at late times.

Within this framework, classes of models with symmetries will be 
represented by invariant subsets of $\mathcal{S}$, the most important 
being the subset $\mathcal{S}_{\rm FL}$ that describes the FL cosmologies.
One expects that the orbits of close-to-FL models will shadow orbits in
$\mathcal{S}_{\rm FL}$.

The appropriate mathematical vehicle for implementing the above program is 
the orthonormal frame formalism%
\footnote{This formalism was introduced in relativistic cosmology by Ellis 
1967, among others, and an extended version was given by van Elst \& Uggla 
1997.},
since it expresses the EFE directly as first order (in time) autonomous 
evolution equations.
It is also necessary to introduce a process of normalization in order to 
create bounded variables.
A choice that has proven effective is the so-called 
\emph{Hubble normalization}. 
This approach to the study of cosmological dynamics has been discussed in 
some detail in Wainwright \& Ellis 1997%
\footnote{We shall abbreviate this reference to WE in this paper.},
and more recently by Coley 2003,
with emphasis on spatially homogeneous models.
Since then the approach has been 
extended to models without symmetry by Uggla \etal 2003.

The plan of this paper is as follows.
In Section~\ref{sec_Hub} we describe the principal features of the 
Hubble-normalized state space, 
and in Section~\ref{sec_c_i} we discuss close-to-FL models and the related 
notion of isotropization.
In Section~\ref{sec_SH} we give an overview of the dynamics of a special 
class of models in the cosmological hierarchy, the nontilted spatially 
homogeneous cosmologies.
In Section~\ref{sec_vac_num} we describe a candidate for the past 
attractor in the Hubble-normalized state space, and mention some recent 
numerical simulations which support this proposal.
In Section~\ref{sec_col} we briefly introduce a modification of 
Hubble normalization which provides a description of nontilted spatially 
homogeneous models that undergo recollapse.
Section~\ref{sec_conclusion} contains our concluding remarks.

\section{The Hubble-normalized state space}\lb{sec_Hub}

In this section we give the motivation for introducing Hubble-normalized 
variables, and describe the principal features of the resulting 
Hubble-normalized state space and evolution equations.

\subsection{Hubble-normalized variables}

One of the rationals for using dynamical systems methods in cosmology is 
the hope that one can describe the evolution of cosmological models near 
the initial singularity by means of a \emph{past attractor} of a 
cosmological dynamical system, and the dynamics at late times in a 
particular epoch by means of a \emph{future attractor}.

In order to formulate the Einstein field equations as a dynamical system 
it is clear that one has to normalize the variables, since near the 
initial singularity physical variables typically diverge and at late times 
typically tend to zero.
Physical considerations suggest that in a cosmological setting, 
normalization with the Hubble scalar of the fundamental congruence is an 
appropriate choice.
Firstly, consider the density parameter $\Omega_m$, which plays a 
fundamental role in cosmology, in that it measures the influence of the 
matter on the dynamics.
In geometrized units, it is defined to be the matter density $\rho$ 
divided 
by $3H^2$, where $H$ is the Hubble scalar.
Secondly, the extent to which the overall expansion of the universe is 
close to isotropy is measured by the ratio of the shear (the trace-free 
part of the expansion tensor) to the Hubble scalar.

The importance of Hubble normalization appears to have first been 
mentioned in the literature by Kristian \& Sachs 1966 (see page 398) in 
connection with their general analysis of the potential constraints that 
observations of distant galaxies can place on the geometry of spacetime.
The role of Hubble-normalized variables emerges again when one analyzes 
the potential constraints that arise from observations of the CMB 
(Maartens \etal 1995, 1996).
In both analyses one finds that bounds are placed on Hubble-normalized 
physical and geometrical quantities.
Our principal interest, however, lies in using Hubble-normalized 
variables as dynamical variables,
which dates back to Collins 1971, who gave a 
qualitative analysis of the dynamics of some special classes of 
cosmological models.

Working within the orthonormal frame formalism we introduce 
Hubble-normalized variables and write the EFE as first order evolution 
equations in the Hubble-normalized state space.
For simple classes of ever-expanding models, for example open FL models, 
and spatially homogeneous models of Bianchi type I, the Hubble-normalized 
state space is bounded. Near the singularity physical variables
such as the matter density diverge, but the Hubble scalar also diverges 
and at such a rate that the corresponding Hubble-normalized quantity, the 
density parameter $\Om_m = \dfrac{\rho}{3H^2}$, remains bounded.
In contrast, if $H$ is close to zero,
then the other physical variables are equally close to zero, 
so that the Hubble-normalized quantities are bounded.

For more general models, however, it turns out that this simple picture is 
not valid -- \emph{the Hubble-normalized state space is unbounded}. 
The primary reason is that 
the Hubble-normalized Weyl curvature tensor can 
assume arbitrarily large values.
All available evidence, however, suggests that 
\emph{the Hubble-normalized variables are bounded into the past}, 
i.e.~on approach to the initial singularity.
In addition, if there is a positive cosmological constant there is strong 
evidence that the Hubble-normalized variables are also bounded into the 
future.
Thus, even though the Hubble-normalized state space is unbounded, one 
expects that the evolution equations will admit a past attractor and a 
future attractor.

\subsection{Asymptotic regimes}

When one formulates the EFE as a dynamical system, one uses a time 
variable $t$ that potentially assumes all real values.
The \emph{asymptotic regimes} are then defined by the limits 
$t \rightarrow -\infty$
and
$t \rightarrow +\infty$.
Cosmologists model the physical universe as a sequence of epochs in time, 
an epoch being identified by which source term is dynamically dominant.
A typical succession of epochs is
\vskip2mm
i) inflationary, dominated by a scalar field $\varphi$,

ii) radiation-dominated, $p=\tfrac13\rho$,

iii) matter-dominated, $p=0$,

iv) accelerating, dominated by a cosmological constant $\Lambda>0$.
%
\vskip2mm
\noindent
With each source term is associated a Hubble-normalized energy density,
\[
	\Om_\varphi,\ \Om_r,\ \Om_m,\ \Om_\Lambda,
\]
and a particular epoch is defined as the time interval during which a 
particular $\Om$ is dominant.

In analytic work one typically assumes the presence of one or two source 
terms%
\footnote{A cosmological constant can be incorporated with little increase 
in complexity in the evolution equations.}
and one investigates the asymptotic regimes,
which are described mathematically by the past attractor and future 
attractor of the dynamical system.
Thus, for example, the past asymptotic regime of the radiation-dominated 
epoch would physically coincide with the end of the inflationary epoch and 
the future asymptotic regime would coincide with the beginning of the 
matter-dominated epoch.

\subsection{The gravitational and matter variables}

A cosmological model is a spacetime with a preferred timelike vector field 
$\mathbf{u}$ whose metric is a solution of the EFE with appropriate 
matter/energy content.
It is assumed that there is an epoch during which the model is expanding, 
i.e.~the preferred vector field satisfies
\[
	H := \tfrac13 \nabla_a u^a >0,
\]
where $H$ is referred to as the Hubble scalar.
We assume furthermore that $\mathbf{u}$ is the normal vector field to a 
family of spacelike hypersurfaces $t= \mbox{constant}$, thereby defining a 
time variable.

We introduce an orthonormal frame $\{ \mathbf{e}_a \}$, with $\mathbf{e}_0 
= \mathbf{u}$, and choose the spatial frame vectors $\mathbf{e}_\alpha$ to 
be Fermi-propagated.
Within this framework, the gravitational field variables are the 
commutation functions $\gam^c{}_{ab}$ associated with the frame, defined 
by
\[
	[ \mathbf{e}_a,\mathbf{e}_b] = \gam^c{}_{ab}\, \mathbf{e}_c.
\]
They may conveniently be expressed in terms of geometric quantities
\be
\lb{geo}
	\{ H,\,\sigma_{\alpha\beta},\,\dot{u}_\alpha,\,
		a_\alpha,\,n_{\alpha\beta}
	\},
\ee
according to
\[
	\gam^\alpha{}_{0\beta} = - 
		\sigma_\beta{}^\alpha-H\d_\beta{}^\alpha,\quad
	\gam^0{}_{0\alpha} = \dot{u}_\alpha,\quad
	\gam^\mu{}_{\alpha\beta} = \eps_{\alpha\beta\nu}\, n^{\mu\nu}
		+ a_\alpha\,\d_\beta{}^\mu - a_\beta\,\d_\alpha{}^\mu,
\]
(see for example WE, page 32). 
Here $\dot{u}_\alpha$ is the acceleration and $\sigma_{\alpha\beta}$ is 
the shear of the congruence $\mathbf{u}$, while $a_\alpha$ and 
$n_{\alpha\beta}$ determine the curvature of the spacelike hypersurfaces 
(WE, pages 18, 19, 34).

As regards matter/energy content we consider a cosmological constant 
$\Lambda$ and a perfect fluid with 4-velocity vector field 
$\tilde{\mathbf{u}}$ and linear barotropic equation of state 
$\tilde{p} =(\gam-1) \tilde{\rho}$, with $1 \leq \gam \leq 2$.
In the orthonormal frame formalism the perfect fluid is described by its 
energy density $\rho$ relative to $\mathbf{e}_0$ and its velocity 
$v^\alpha$ relative to $\mathbf{e}_0$,
i.e.~the projection of $\tilde{\mathbf{u}}$ orthogonal to $\mathbf{e}_0$
(see Uggla \etal 2003, equations (2.1)--(2.6) for details).
So the gravitational variables~(\ref{geo}) are augmented by the matter
variables
\be
	\{ \rho,\,v_\alpha,\,\Lambda\}.
\ee
The velocity $v^\alpha$ is dimensionless and satisfies
\[
	0 \leq v_\alpha v^\alpha <1.
\]
We now introduce Hubble-normalized variables%
\footnote{Hubble-normalized variables have also been defined for 
cosmological models with source terms other than a perfect fluid with 
linear equation of state and a cosmological constant,
e.g. magnetic fields (LeBlanc 1998), scalar fields (Coley \& van den 
Hoogen 2000)
 and brane matter fields (Coley etal 2004).}
 and differential operators 
according to
\begin{gather}
\lb{hubnormvar}
	\{ \Sigma_{\alpha\beta},\,\dot{U}_\alpha,\,
                A_\alpha,\,N_{\alpha\beta} \} :=
	\{ \sigma_{\alpha\beta},\,\dot{u}_\alpha,\,
                a_\alpha,\,n_{\alpha\beta} \} /H,\quad
	\{ \Om_m,\,\Om_\Lambda \} := \{\rho,\,\Lambda\}/(3H^2),
\\
	\parb_0 = \frac1H \mathbf{e}_0,\quad
	\parb_\alpha = \frac1H \mathbf{e}_\alpha.
\notag
\end{gather}
It is also necessary to introduce the derivatives of the normalization 
factor, namely the 
\emph{deceleration parameter} $q$ and the 
\emph{spatial Hubble gradient} $r_\alpha$, defined by
\be
	q+1 := - \frac1H \parb_0 H,\quad
	r_\alpha := - \frac1H \parb_\alpha H.
\ee
At this stage we introduce the separable volume gauge (Uggla \etal 2003), 
which is characterized by
\[
	\dot{U}_\alpha = r_\alpha
\]
and the existence of local coordinates $t$, $x^i$ such that
\[
	\parb_0 = \ptl_t\,,\quad
	\parb_\alpha = E_\alpha{}^i \ptl_i\,,
\]
with the preferred hypersurfaces being given by $t=\mbox{constant}$.
Here $\ptl_t$ and $\ptl_i$ denote partial differentiation with respect to 
$t$ and $x^i$.
The deceleration parameter $q$ can then be expressed in terms of the
variables in (\ref{hubnormvar}) and the differential operator 
$\parb_\alpha$, 
using the Raychaudhuri equation.

With the above choices of gauge the Hubble-normalized state vector becomes
\[
	\mathbf{X} = ( E_\alpha{}^i,\,r_\alpha,\
	\Sigma_{\alpha\beta},\,A_\alpha,\,N_{\alpha\beta},\
	\Om_m,\,\Om_\Lambda,\,v_\alpha )^T.
\]
The EFE, the Jacobi identities and the commutators lead to a system of 
evolution equations and a system of constraints for the components of 
$\mathbf{X}$.
In order to describe the structure of these equations, it is convenient to 
decompose the state vector $\mathbf{X}$ as follows:
\be
	\mathbf{X} = ( E_\alpha{}^i,\,r_\alpha)^T \oplus \mathbf{Y},
\ee
where
\be
\lb{Y}
	\mathbf{Y} = ( \Sigma_{\alpha\beta},\,A_\alpha,\,N_{\alpha\beta},\
        \Om_m,\,\Om_\Lambda,\,v_\alpha )^T.
\ee
The evolution equations have the following form:
\begin{align}
\lb{bsys}
	\ptl_t E_\alpha{}^i &= 
	(q\,\d_\alpha{}^\beta-\Sig_\alpha{}^\beta)E_\beta{}^i
\\
\lb{evor}
	\ptl_t r_\alpha &=
	(q\,\d_\alpha{}^\beta-\Sig_\alpha{}^\beta)\,r_\beta 
	+E_\alpha{}^i \ptl_i q
\\
\lb{evoY}
	\ptl_t Y_A &=
	F_A(\mathbf{Y}) 
	+ F_A{}^{B\alpha}(\mathbf{Y}) E_\alpha{}^i\ptl_i Y_B
	+ F_A{}^{\alpha\beta}(\mathbf{Y})E_\alpha{}^i\ptl_i r_\beta
	+ F_A{}^\alpha(\mathbf{Y})\, r_\alpha\,,
\end{align}
where the deceleration parameter $q$ is given by
\be
\lb{q}
	q = 2 \Sig^2 + \tfrac12 
	\frac{[(3\gam-2)+(2-\gam)v^2]}{1+(\gam-1)v^2}\Om_m
	-\Om_\Lambda - \tfrac13(E_\alpha{}^i \ptl_i - 
	2A_\alpha)\,r^\alpha.
\ee
There is also a set of constraints that can be written symbolically as
\be
\lb{esys}
	0 = \mathcal{C}(\mathbf{X},\,\ptl_i\mathbf{X}).
\ee
One of the constraints, the so-called Gauss constraint, is of particular 
importance, and we thus give it specifically%
\footnote{Lim \etal 2004, equation (2.27).}:
\be
\lb{CGauss}
	\Om_m + \Om_\Lambda + \Om_k + \Sig^2 = 1,
\ee
where $\Sig^2$ is the Hubble-normalized shear scalar given by
\be
\lb{Sig_sq}
	\Sig^2 = \tfrac16 \Sig_{\alpha\beta} \Sig^{\alpha\beta},
\ee
and $\Om_k$ is the Hubble-normalized spatial curvature%
\footnote{$\Om_k$ is defined by $\Om_k = - \frac{ {}^3\!R}{6H^2}$, where
${}^3\!R$ is the curvature scalar of the metric induced on the
hypersurfaces $t=\mbox{constant}$.}
of the hypersurfaces $t=\mbox{constant}$, given by
\be
	\Om_k = -\tfrac23(\parb_\alpha-r_\alpha) A^\alpha
		+\tfrac16 N_{\alpha\beta} N^{\alpha\beta}
		-\tfrac1{12}(N_\alpha{}^\alpha)^2
		+A_\alpha A^\alpha.
\ee

The detailed form of equations (\ref{bsys})--(\ref{esys}) 
 is given in Lim \etal 2004 (equations 
(2.18)--(2.30) with $R^\alpha=0$).
What concerns us here is the overall structure of this system.
We make a number of observations.
\begin{itemize}
\item[i)]
	The equations are first order evolution equations in time, but do 
	contain second order spatial derivatives, namely the second 
	derivatives of the spatial Hubble gradient 
	$\ptl_i \ptl_j r_\alpha$.
	These derivatives appear in (\ref{evor}), on account of (\ref{q}).
	In this respect, the equations are reminiscent of a system of 
	quasi-linear diffusion equations.
	We shall discuss this matter further in Section~\ref{sec_vac_num}.

\item[ii)]
	The frame variables $E_\alpha{}^i$ enter into the remaining
equations only through the spatial differential operator $\parb_\alpha$.
Recalling that $r_\alpha$ is the spatial Hubble gradient we thus regard
the variables $E_\alpha{}^i$ and $r_\alpha$ as controlling the spatial
inhomogeneity in the cosmological model, while the variables in
$\mathbf{Y}$ directly determine the spacetime geometry (i.e.~the 
gravitational field) and the matter content.

\end{itemize}

\subsection{The cosmological hierarchy}\lb{sec2.3}

We consider three classes of ever-expanding cosmological models, namely, 
Friedmann-Lema\^{\i}tre (FL) models, spatially homogeneous (SH) models and 
general models, i.e.~models without symmetry.
The SH models admit a three-parameter local group $G_3$ of isometries 
acting on spacelike hypersurfaces, 
while in the FL models the isometry group 
is a $G_6$. For brevity we shall refer to the models without symmetry as 
\emph{$G_0$ cosmologies}.
So we have the following cosmological hierarchy:
\[
	{\rm FL} \subset {\rm SH} \subset G_0.
\]
We will use $\H$ to denote the corresponding Hubble-normalized state 
spaces:
\[
	\H_{\rm FL} \subset \H_{\rm SH} \subset \H_{G_0}.
\]

One can construct a more detailed hierarchy by including models with two 
or one spacelike Killing vectors.
Classes of models with two Killing vectors are variously referred to%
\footnote{See Uggla \etal 2003, Table 3, for selected references.}
as Gowdy%
\footnote{See Andersson \etal 2004 for a detailed analysis of Gowdy models 
using Hubble-normalized variables.},
$T_2$-symmetric or $G_2$ cosmologies.
Models with one Killing vector are referred to as $U(1)$-symmetric%
\footnote{See Berger 2004 for a discussion of Mixmaster dynamics 
in these models.}
or $G_1$ cosmologies.

\subsubsection*{SH cosmologies}

The SH cosmologies are obtained by requiring that the spatial frame 
derivatives of the gravitational field and matter variables $\mathbf{Y}$, 
and of the normalization factor $H$ be zero, i.e.
\be
\lb{SHres}
	\parb_\alpha \mathbf{Y}=0,\quad r_\alpha=0.
\ee
It then follows that all the dimensional commutation functions and matter 
variables are constant on the hypersurface $t=\mbox{constant}$, which are 
thus the orbits of a three-parameter group $G_3$ of isometries.
The evolution equations (\ref{evor}) and (\ref{evoY}) imply that the SH 
restrictions (\ref{SHres}) define an invariant set of the full evolution 
equations, which we shall call the \emph{SH invariant set}.
Indeed, equation (\ref{evor}) is trivially satisfied, and equation 
(\ref{evoY}) reduces to a system of ordinary differential equations, 
namely
\be
\lb{SH1}
	\ptl_t Y_A = F_A(\mathbf{Y}).
\ee
The nontrivial constraints become purely algebraic restrictions on 
$\mathbf{Y}$, which we write symbolically as
\be
\lb{SH2}
	\mathcal{C}(\mathbf{Y})=0.
\ee
An important consequence of this specialization is that the 
evolution equation (\ref{bsys}) for $E_\alpha{}^i$ decouples from the 
evolution equation for $\mathbf{Y}$, which means that
\emph{the dynamics of SH cosmologies can be analyzed using only
equations (\ref{SH1}) and (\ref{SH2})}. In this context, one can think of 
the variable $\mathbf{Y}$ in equation (\ref{Y}) as defining a
\emph{reduced Hubble-normalized state space},
of finite dimension, for the SH cosmologies.
The Bianchi classification of the isometry group then leads to a hierarchy 
of invariant subsets.

In the SH context the restriction%
\footnote{This restriction means that the fluid 4-velocity is orthogonal 
to the orbits of the group of isometries (Ellis \& MacCallum 1969).}
 $v^\alpha=0$ defines an invariant 
subset, giving the so-called \emph{nontilted SH cosmologies}.
This class of models have been analyzed in detail in the literature.
We give an overview of the known results in Section~\ref{sec_SH}.
We note in passing that tilted SH cosmologies (i.e. $v^\alpha\neq0$) were 
first studied by King \& Ellis 1973. Considerable progress has been made 
recently, using Hubble-normalized variables 
(see Hewitt \etal 2001, Hervik 2004, Coley \& Hervik 2004).

\subsubsection*{FL cosmologies}

The FL cosmologies are obtained by imposing the restrictions
\[
	\Sig_{\alpha\beta}=0,\quad
	v_\alpha=0,\quad
	r_\alpha=0,
\]
which is equivalent to requiring that the fluid has zero shear, vorticity 
and acceleration%
\footnote{Refer to WE, Section 2.4 for the characterizations of FL.}.
It follows that the density parameters are constant on the hypersurface
$t=\mbox{constant}$%
\footnote{The first follows from the evolution equation for $v_\alpha$ and 
the second from the definition of $\Om_\Lambda$ and $r_\alpha$.}:
\[
	\parb_\alpha \Om_m=0,\quad
	\parb_\alpha \Om_\Lambda=0.
\]
In addition%
\footnote{See Uggla \etal 2003, equations (A10), (A11), (A14) and (A15).}
the Weyl curvature is zero,
\[
	\mathcal{E}_{\alpha\beta} = 0 = \mathcal{H}_{\alpha\beta},
\]
and the 3-Ricci curvature satisfies
\be
\lb{Sab0}
	\mathcal{S}_{\alpha\beta}=0,\quad
	\parb_\alpha \Om_k =0.
\ee

The dynamics of the FL models is governed by the matter evolution 
equations which simplify to a system of two ODEs,
\begin{align}
\lb{SH_Om}
	\ptl_t \Om_m &= [2q-(3\gam-2)]\Om_m
\\
\lb{SH_OL}
	\ptl_t \Om_\Lambda &= 2(q+1) \Om_\Lambda,
\end{align}
with
\be
\lb{SH_q}
	q = \tfrac12(3\gam-2)\Om_m -\Om_\Lambda.
\ee
The Gauss constraint (\ref{CGauss}) simplifies to
\be
\lb{SH_CGauss}
	        \Om_m + \Om_k + \Om_\Lambda = 1,
\ee
where all three quantities are functions of time only.
For future reference we note that the curvature parameter $\Om_k$ 
satisfies the evolution equation
\be
\lb{SH_Omk}
	\ptl_t \Om_k = 2q \Om_k,
\ee
as follows from (\ref{SH_Om}), (\ref{SH_OL}) and (\ref{SH_CGauss}).

The remaining evolution equations simplify to
\[
	\ptl_t E_\alpha{}^i = q E_\alpha{}^i,\quad
	\ptl_t A_\alpha = q A_\alpha,\quad
	\ptl_t N_{\alpha\beta} = q N_{\alpha\beta},
\]
and hence decouple from (\ref{SH_Om}) and (\ref{SH_OL}).
These variables reflect the choice of spatial frame, and do not affect the 
essential dynamics.
Since $q=q(t)$, these equations can be integrated to yield
\[
	E_\alpha{}^i = f(t) \hat{E}_\alpha{}^i,\quad
	A_\alpha = f(t) \hat{A}_\alpha,\quad
	N_{\alpha\beta} = f(t) \hat{N}_{\alpha\beta},
\]
where hatted variables depend only on the spatial coordinates $x^i$, and
$f'(t) = q f(t)$.
These quantities are restricted be equation (\ref{Sab0}).
Models with negative, zero and positive spatial curvature are 
characterized respectively by
\[
	\Om_k >0,\quad
	\Om_k=0,\quad
	\Om_k<0,
\]
and in the corresponding canonical spatial frames we have
\begin{gather*}
	N_{\alpha\beta}=0,\quad A_\alpha=(A,0,0),\quad \Om_k = A^2,
\\
	N_{\alpha\beta}=0,\quad A_\alpha=0,\quad \Om_k=0,
\\
	N_{\alpha\beta}=\text{diag}(N,N,N),\quad A_\alpha=0,\quad 
	\Om_k = -\tfrac14 N^2.
\end{gather*}
Thus, the FL cosmologies with perfect fluid and cosmological constant are 
described by a reduced Hubble-normalized state space, namely the 
two-dimensional space with the state vector $(\Om_m,\ \Om_\Lambda)$, and 
evolution equations (\ref{SH_Om})--(\ref{SH_q}).
It follows from (\ref{SH_CGauss}) that for models with negative and zero 
spatial curvature ($\Om_k\geq0$), the state space is bounded.
This state space is shown in Figure~\ref{fig3.1} in Section~\ref{sec_c_i}, 
using $\Om_k$ and $\Om_\Lambda$ as variables.

\subsection{Cosmological equilibrium points}

In the analysis of the dynamics of SH models using the reduced 
Hubble-normalized state space $\H_{\rm SH}$, the equilibrium points
(i.e.~fixed points) of the dynamical system, defined by
\[
	\ptl_t \mathbf{Y}=0,
\]
naturally play a significant role.
At this stage the Kasner vacuum solutions
come into play, given in Hubble-normalized variables by
\be
\lb{K_vac}
	N_{\alpha\beta}=0,\quad A_\alpha=0,\quad \Sig^2=1,\quad
	\Om_m=0,\quad \Om_\Lambda=0,\quad q=2,
\ee
with
\[
	\ptl_t \Sig_{\alpha\beta}=0.
\]
These equilibrium points%
\footnote{The fact that the Kasner invariant set, defined by
(\ref{K_vac}), consists only of equilibrium points is a consequence of
the fact that we are using a Fermi-propagated spatial frame. See Uggla
\etal 2003, page 7, for a different spatial gauge choice, which leads to a
circle of equilibrium points, and ``frame transition orbits".}
 form a 4-sphere (recall the definition 
(\ref{Sig_sq}), and that $\Sig_{\alpha\beta}$ 
is trace-free), which we shall call the \emph{Kasner sphere} 
$\mathcal{K}$.
If $\Sig_{\alpha\beta}$ is diagonal, then $\mathcal{K}$ reduces to a 
circle.

Two other important equilibrium points lie in the FL invariant set and 
satisfy
\[
	N_{\alpha\beta}=0,\quad A_\alpha=0,\quad \Sig_{\alpha\beta}=0,
	\quad
	v_\alpha=0,
\]
and one of the following sets of conditions:
\footnote{The de Sitter solution has other representations that involve 
the peculiar velocity $v_\alpha$. See Table~\ref{tab1}.}
\begin{alignat}{5}
\lb{FL}
	&\text{flat FL}\qquad & \Om_m &=1,\quad& 
	\Om_\Lambda&=0,\quad& q&=\tfrac12(3\gam-2)
\\
\lb{dS}
	&\text{de Sitter}\qquad & \Om_m &=0,\quad& 
        \Om_\Lambda&=1,\quad& q&=-1.
\end{alignat}
The local stability of the flat FL and de Sitter equilibrium points is of 
importance in connection with the phenomenon of isotropization, as 
discussed in Section~\ref{sec_iso}.
In the 
following tables we give the dimensions of the stable and unstable 
manifolds of these equilibrium points, for both nontilted and tilted SH 
models.

\begin{table}[ht]
\caption{Stability of the flat FL and de Sitter equilibrium points in the
state space for nontilted SH models (six dimensional)}\lb{tab_dim1}
\centering{
\begin{spacing}{1.1}
\begin{tabular}{ccc}
\\
Equilibrium&	dimension of the	&       dimension of the
\\
point	&	stable manifold		& 	unstable manifold
\\
\hline
flat FL & 2 & 4
\\
de Sitter & 6 & 0
\\
\hline
\end{tabular}\end{spacing}
}
\end{table}

\begin{table}[ht]
\caption{Stability of the flat FL and de Sitter equilibrium points in the
state space for tilted SH models (nine dimensional)}
\centering{
\begin{spacing}{1.1}
\begin{tabular}{ccc}
\\
Equilibrium&       dimension of the
&       dimension of the
\\
point&       stable manifold
&       unstable manifold
\\
\hline
flat FL & 5 & 4
\\
de Sitter, $0<\gam<\tfrac43$ & 9 & 0
\\
de Sitter, $\tfrac43<\gam<2$ & 6 & 3
\\
\hline
\end{tabular}\end{spacing}
}
\end{table}

The SH equilibrium points, with the exception of de Sitter, represent 
self-similar solutions of the EFE,
admitting a four-parameter group $H_4$ of similarities.
We refer to WE, Section 9.1, for a complete list of these self-similar 
solutions within the class of nontilted SH models.

\subsection{The silent boundary}\lb{sec_silent_b}

        On account of (\ref{bsys}) and (\ref{evor}), the conditions
\be
\lb{silent_cond}
        E_\alpha{}^i=0,\quad r_\alpha=0
\ee
define an invariant set,
the so-called \emph{silent boundary}%
\footnote{This notion was first introduced in van Elst \etal 2002 and
Uggla \etal 2003. See Andersson \etal 2004 for further discussion.},
that forms part of the boundary of the Hubble-normalized
state space.  
Solutions of the evolution equations in this invariant set do not
correspond to solutions of the EFE, however,
since the frame variables have to satisfy $\det(E_\alpha{}^i)\neq0$.
Nevertheless, it appears that this invariant set plays a fundamental role
in describing the asymptotic behaviour of cosmological models.
Indeed current investigations suggest that the past attractor
 and the future attractor for $G_0$ cosmologies are subsets of the above
invariant set (subject to $\Lambda>0$ in the case of the future
attractor). We will discuss this matter further in Sections~\ref{sec_c_i}
and~\ref{sec_vac_num}.
The condition $E_\alpha{}^i=0$
in the definition of the silent boundary
 implies that $\parb_\alpha \mathbf{Y}=0$.
It thus follows that the evolution equations and constraints, when 
specialized to the silent boundary, are precisely the SH evolution 
equations and constraints, as given by (\ref{SH1}) and (\ref{SH2}).
In addition each self-similar $H_4$ solution determines an equilibrium 
point in the silent boundary.

\section{Close-to-FL models and isotropization}\lb{sec_c_i}

\subsection{Close-to-FL models}\lb{sec_close_to__FL}

It is customary to regard a cosmological model as being ``close to FL" if 
its metric is a linear perturbation of an FL metric (e.g.~Bardeen 1980), 
i.e.~the dynamics of such models are governed by the \emph{linearized} 
EFE.%
\footnote{For a fully covariant approach to scalar perturbation of FL, 
which physically describe density fluctuations, we refer to Ellis \& Bruni 
1989, and Ellis \etal 1989.}
An alternative approach is to regard a cosmological model as being 
``close to FL" if an appropriate set of Hubble-normalized anisotropy 
parameters are small.%
\footnote{See WE pages 63-4, for a tentative definition of ``close-to-FL".
Kristian \& Sachs 1966 give a list of parameters, called ``adjustable 
parameters" in their Table 1 (page 393) and referred to as
``anisotropy and inhomogeneity parameters" on page 398.}
For simplicity we will restrict our attention to two parameters. Firstly 
we consider the shear parameter $\Sig$, defined by (\ref{Sig_sq}), which 
describes the anisotropy in the rate of expansion of the fundamental 
congruence. Secondly we consider the Weyl parameter $\Weyl$ defined by
\[
	\Weyl^2 = \frac{1}{6H^4} 
		(E_{\alpha\beta} E^{\alpha\beta}
		+H_{\alpha\beta} H^{\alpha\beta}),
\]
where $E_{\alpha\beta}$ and $H_{\alpha\beta}$ are the 
electric and magnetic parts of the Weyl curvature tensor relative to the 
fundamental congruence.
The Weyl parameter can be thought of as quantifying the intrinsic 
anisotropy of the gravitational field.

As described in Section~\ref{sec2.3},
the FL cosmologies with negative or zero spatial curvature and positive 
cosmological constant have a two-dimensional Hubble-normalized state 
space.
Using $\Om_k$ and $\Om_\Lambda$ as independent variables, the state space 
is shown in Figure~\ref{fig3.1}.
The equilibrium points represent the flat FL model (F), de Sitter 
model (dS), and the Milne model (M).
One sees that F is the past attractor and dS is the future attractor. The
Milne point M is a saddle, which means that a subset of open FL models  
with $\Lambda>0$ will have a ``close-to-Milne" epoch of finite duration.

\begin{figure}[ht]
  \begin{center}
    \epsfig{file=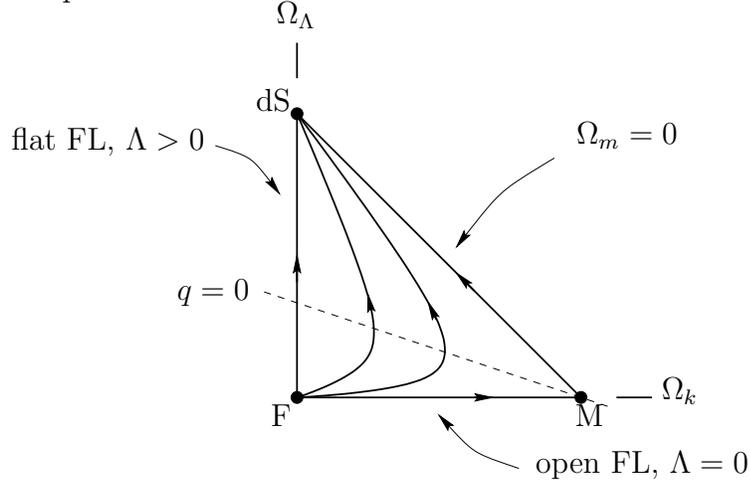,width=3in}
\setlength{\unitlength}{1mm}
\begin{picture}(120,0)(0,0)
\put(41,64){\makebox(0,0)[b]{$\Om_\Lambda$}}
\put(85,48){\makebox(0,0)[b]{$\Om_m=0$}}
\put(16,47){\makebox(0,0)[b]{flat FL, $\Lambda>0$}}
\put(87,4){\makebox(0,0)[b]{open FL, $\Lambda=0$}}          
\put(39,11){\makebox(0,0)[b]{F}}
\put(30,27){\makebox(0,0)[b]{$q=0$}}
\put(38,53){\makebox(0,0)[b]{dS}}
\put(80,11){\makebox(0,0)[b]{M}}
\put(92,14){\makebox(0,0)[b]{$\Om_k$}}
\end{picture}
\caption{The Hubble-normalized state space for flat and open FL 
	cosmologies with $\Lambda>0$ 
	and equation of state 
	$p=(\gam-1)\rho$, $\tfrac23<\gam<2$.}\lb{fig3.1} 
\end{center}
\end{figure}

When one regards the FL models as a subset of the SH models, a 
complication arises:
a specific FL model is represented by infinitely many orbits in the 
reduced Hubble-normalized SH state space.
This multiple representation occurs because a flat FL model admits a $G_3$ 
of Bianchi type I and one of Bianchi type VII$_0$, while an open FL model 
admits a $G_3$ of Bianchi type V and one of Bianchi type VII$_h$, for any 
$h>0$ (see Ellis \& MacCallum 1969).
The state space can be described by the variables
\[
	(\Om_\Lambda,\ \Om_k,\ N),
\]
where
\[
	\Om_k=A^2=h N^2.
\]
Here $h$ is the group parameter and
 $A$ and $N$ are spatial curvature variables defined by
\[
	N_{\alpha\beta} = \text{diag}(0,N,N),\quad
	A_\alpha = (A,0,0).
\]
The state space, which we will denote by $\H_{\rm FL}$, is shown 
in Figure~\ref{fig3.2}.

\begin{figure}[ht]
  \begin{center}  
    \epsfig{file=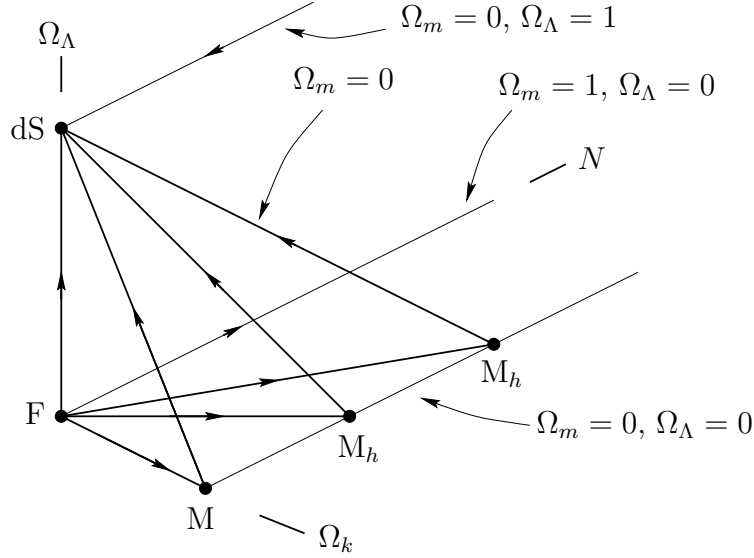,width=4.5in}
\setlength{\unitlength}{1mm}
\begin{picture}(170,0)(0,0)
\put(42,70){\makebox(0,0)[b]{$\Om_\Lambda$}}
\put(115,63){\makebox(0,0)[b]{$\Om_m=1$, $\Om_\Lambda=0$}}
\put(102,72){\makebox(0,0)[b]{$\Om_m=0$, $\Om_\Lambda=1$}}
\put(80,64){\makebox(0,0)[b]{$\Om_m=0$}}
\put(120,18){\makebox(0,0)[b]{$\Om_m=0$, $\Om_\Lambda=0$}}
\put(79,3){\makebox(0,0)[b]{$\Om_k$}}
\put(39,20){\makebox(0,0)[b]{F}}
\put(38,58){\makebox(0,0)[b]{dS}}
\put(113,54){\makebox(0,0)[b]{$N$}}
\put(82,15){\makebox(0,0)[b]{M$_h$}}
\put(61,6){\makebox(0,0)[b]{M}}
\put(101,25){\makebox(0,0)[b]{M$_h$}}
\end{picture}
\end{center}
\caption{The Hubble-normalized state space for flat and open 
FL cosmologies with $\Lambda>0$ and 
equation of state $p=(\gam-1)\rho$, $\tfrac23 
< \gam <2$, showing multiple representations.}\lb{fig3.2} 
\end{figure}

\begin{figure}[ht]
  \begin{center}
    \epsfig{file=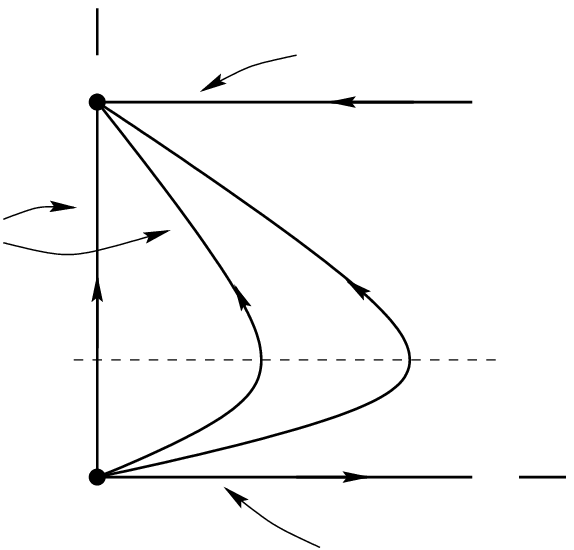,width=2.3in} 
\setlength{\unitlength}{1mm}
\begin{picture}(120,0)(0,0)
\put(41,62){\makebox(0,0)[b]{$\Om_\Lambda$}}
\put(38,10){\makebox(0,0)[b]{F}}
\put(79,3){\makebox(0,0)[b]{flat FL, $\Lambda=0$}}
\put(70,55){\makebox(0,0)[b]{de Sitter}}
\put(33,23){\makebox(0,0)[b]{$q=0$}}
\put(17,36){\makebox(0,0)[b]{flat FL, $\Lambda>0$}}
\put(37,50){\makebox(0,0)[b]{dS}}
\put(93,11){\makebox(0,0)[b]{$N$}}
\end{picture}
\end{center}
\caption{The state space $\Om_k=0$, showing multiple 
representations of the flat FL models.}\lb{fig3.3}
\end{figure}

We see that the state space is the union of a one-parameter family of 
``leaves",
and that the Milne model is represented by a line of equilibrium points 
${\rm M}_h$, one in each leaf.
The family of orbits in one leaf with $h>0$ are physically equivalent to 
those in another. The orbits in the leaf with $h=0$, and hence $\Om_k=0$,
 are multiple 
representations of flat FL models with $\Om_\Lambda=0$ or $\Om_\Lambda>0$.
In particular, the orbit with $\Om_\Lambda=0$, $\Om_k=0$ is an unbounded 
orbit that describes the flat FL model,
and the orbit with $\Om_k=0$, $\Om_m=0$ is an unbounded orbit that 
describes the de Sitter model.
The state space $\Om_k=0$ is shown in Figure~\ref{fig3.3}.

The FL state space $\H_{\rm FL}$ is an invariant subset of the SH state 
space $\H_{\rm SH}$, defined by the requirement that the shear of the 
congruence normal to the $G_3$ orbits is zero.
It is natural to regard an SH model as being close to FL during some epoch 
if its orbit lies in a sufficiently small neighbourhood of 
the FL state space $\H_{\rm FL}$.
The Hubble-normalized Weyl tensor is unbounded in such a neighbourhood, 
however, with the result that close-to-FL models can have large 
Hubble-normalized Weyl curvature.
In particular there are models with $\Om_m\approx 1$, $\Sig\approx0$ and 
$\Weyl$ large (close to flat FL) and models with
$\Om_\Lambda\approx 1$, $\Sig\approx0$ and    
$\Weyl$ large (close to de Sitter), during some epoch.

It is also of interest to consider the connection with perturbed FL 
models. Are orbits in a neighbourhood of $\H_{\rm FL}$ described 
accurately by solutions of the linearized EFE, or do nonlinear effects 
come into play?
Consideration of SH cosmologies of Bianchi type VII$_0$ shows that
the latter occurs.
Linearizing the Hubble-normalized evolution equations about the FL 
invariant set $\Om_\Lambda=0$ and $\Om_k=0$ gives
\[
	\Weyl \approx C e^{3(\gam-1)\tau},\quad
	\text{for $\tfrac23<\gam<2$,}
\]
while an analysis of the exact equations for large $\tau$ gives
\[
	\Weyl \approx 
	\begin{cases}
		C e^{3(\gam-1)\tau}, &\text{for $\tfrac23<\gam<\tfrac43$}
		\\
		C \tau^{-\frac32}{e^\tau}, &\text{for $\gam=\frac43$}
		\\
		C e^{\frac32(2-\gam)\tau}, &\text{for $\frac43<\gam<2$,}
	\end{cases}
\]
(Wainwright \etal 1999
).
Thus for this class of models, linearization does not give the correct 
dynamics.

It would be of interest to extend this analysis to $G_0$ cosmologies, 
i.e.~consider $\H_{\rm FL}$ as an invariant set of the state space of 
$G_0$ cosmologies and analyze the dynamics of close-to-FL cosmologies 
in this context. It would be necessary to determine all representations of 
the FL models within the $G_0$ state space, which would give an invariant 
set larger than $\H_{\rm FL}$.

\subsection{Isotropization}\lb{sec_iso}

The term ``isotropization" is used in three different contexts in 
cosmology:
\begin{itemize}
\item[i)]	asymptotic isotropization into the future,
\item[ii)]	isotropic initial singularity,
\item[iii)]	intermediate isotropization.
\end{itemize}
We discuss these dynamic phenomena in turn.

\subsubsection*{Asymptotic isotropization into the future}

A well known result of Wald 1983 states that any SH cosmology with zero or 
negative spatial curvature (i.e.~$\Om_k\geq0$) approaches the de Sitter 
solution in the sense that
\[
	\lim_{t\rightarrow +\infty} \Sig =0,\quad
	\lim_{t\rightarrow +\infty} \Om_\Lambda=1.
\]
Wald proved this result without making specific assumptions about the 
matter/energy distribution.
More detailed analyses of SH cosmologies with a tilted perfect fluid 
(Goliath \& Ellis 1999) revealed that a bifurcation occurs at 
$\gam=\tfrac43$:
if $\gam \geq \tfrac43$ the peculiar velocity $v^\alpha$ of the fluid does 
not tend to zero. Specifically,
$\lim\limits_{t\rightarrow +\infty} v_\alpha = c_\alpha$, where 
$c_\alpha c^\alpha \leq 1$ if $\gam = \tfrac43$ and 
$c_\alpha c^\alpha = 1$    if $\tfrac43 < \gam < 2$.
This behaviour occurs for the following reason.
For the de Sitter solution the reduced state vector $\mathbf{Y}$ in 
(\ref{Y}) is given by
\be
\lb{dS_full}
	N_{\alpha\beta}=0,\quad A_\alpha=0,\quad \Sig_{\alpha\beta}=0,\quad
	\Om_m=0,\quad \Om_\Lambda=1,
\ee
with $v_\alpha$ subject to the evolution equation
\[
	\ptl_t v_\alpha = \frac{(3\gam-4)(1-v^2)}{1+(\gam-1)v^2} v_\alpha.
\]
In the reduced Hubble-normalized state space the de Sitter solution is 
thus represented by a set of equilibrium points, given by (\ref{dS_full}) 
and with $v_\alpha$ subject to $v_\alpha=0$, or $v^2=1$ (a point and a 
sphere) if $\gam \neq \tfrac43$, and subject to $v^2 \leq 1$ (a solid 
sphere) if $\gam = \tfrac43$.
The limiting behaviour of $v^\alpha$ determines the future attractor in 
the reduced Hubble-normalized state space, as shown in Table~\ref{tab1}.

\begin{table}[ht]
\caption{The future attractor for SH cosmologies with $\Lambda>0$ and 
equation of state
 \mbox{$p=(\gam-1)\rho$}.}
\lb{tab1}
\centering{
\begin{spacing}{1.5}
\begin{tabular}{cl}
\\
Equation of state	& Future attractor
\\
\hline
$0 < \gam < \tfrac43$	& de Sitter with zero tilt ($v_\alpha=0$)
\\
$\gam = \tfrac43$	& de Sitter with intermediate tilt 
				($0 \leq v_\alpha v^\alpha \leq 1$)
\\
$\tfrac43 < \gam < 2$	& de Sitter with extreme tilt
				($v_\alpha v^\alpha = 1$)
\\
\hline
\end{tabular}\end{spacing}
}
\end{table}

The future attractor for SH cosmologies with $\Lambda>0$ forms the basis 
for the description of $G_0$ cosmologies that are future asymptotic to the 
de Sitter solution, as we now explain.
The frame variables $E_\alpha{}^i$
 satisfy equation (\ref{bsys}), which, 
when specialized to the de Sitter equilibrium solution (\ref{dS_full}) can 
be solved to give
\[
	E_\alpha{}^i =\hat{E}_\alpha{}^i e^{-t},
\]
where the $\hat{E}_\alpha{}^i$ are constants.
It follows that
\be
\lb{dS_silent}
	\lim_{t\rightarrow+\infty} E_\alpha{}^i =0.
\ee
In addition the constraints imply that the spatial Hubble gradient is zero 
($r_\alpha=0$).
Thus in the Hubble-normalized $G_0$ state space, the de Sitter
solution is described by orbits that are future asymptotic to a de Sitter   
equilibrium point on the silent boundary.

$G_0$ cosmologies that are future asymptotic to the de Sitter solution are 
described within the Hubble-normalized state space as follows.
Along a fixed timeline, the evolution is described by an  
orbit that is future asymptotic to a de Sitter equilibrium point on the
silent boundary.
The specific equilibrium point will depend on the equation of state 
parameter $\gam$ according to Table~\ref{tab1}.
The idea is that the evolution is described by an orbit that is attracted
to the silent boundary, i.e. (\ref{dS_silent}) holds,
and within the silent boundary, the evolution is governed by the SH
evolution equations, as described in Section~\ref{sec_silent_b}.
The asymptotic dependence of the Hubble-normalized variables for this 
class of models has been determined by systematically integrating the 
evolution equations (Lim \etal 2004).
The fact that the orbit is asymptotic to the silent boundary means that 
the spatial dependence is asymptotically unrestricted, which manifests 
itself in the appearance of an arbitrary 3-metric in the asymptotic 
expansion of the spacetime metric (Lim \etal 2004).

The issue of how large is the class of $G_0$ cosmologies that are future 
asymptotic to de Sitter has not been completely resolved. For recent 
progress we refer to Rendall 2003.

\subsubsection*{Isotropic singularities}

The second type of asymptotic isotropization, namely isotropization
at the initial singularity, arises in connection with the notion of
{\em quiescent cosmology\/} (cf., e.g., Barrow 1978), which
provides an alternative to cosmic inflation. The idea is that, due
to entropy considerations on a cosmological scale, a suitable
initial condition for the universe is that the Weyl curvature
should be zero (or at least dynamically unimportant) at the initial
singularity; this is the {\em Weyl curvature hypothesis\/}, proposed by  
Penrose 1979 (see page~630). This hypothesis leads to the notion of 
an {\em isotropic initial singularity\/}.  
Recently, Khalatnikov \etal 2003 (see page 3) have suggested that a 
solution with an isotropic singularity could represent an 
\emph{intermediate} asymptotic regime during expansion of the universe 
after inflation.
Cosmological initial
singularities of this type first arose as a special case in the
general analysis of initial singularities performed by Lifshitz \&
Khalatnikov 1963 (see page~203). This type of
initial singularity was also encountered in the work of Eardley 
\etal 1972 (see page~101). Subsequently, 
Goode \& Wainwright~1985 gave a formal definition of an 
isotropic
initial singularity, using a conformally related metric, and
derived various properties. 
Heuristically, the model approaches the flat FL model near the 
singularity:
\[
        \lim_{t\rightarrow -\infty} \Sig =0,\quad
        \lim_{t\rightarrow -\infty} \Om_m  =1.
\]

For SH models, the occurrence of an isotropic singularity is characterized 
by the orbit in the reduced Hubble-normalized state space being past 
asymptotic to 
the flat FL equilibrium point F.
In other words, the orbit of a SH model with an isotropic singularity lies 
in the unstable manifold of the flat FL equilibrium point, which confirms 
that isotropic singularities occur only for a set of initial conditions of 
measure zero.

The above behaviour forms the basis for the description of $G_0$ 
cosmologies that admit an isotropic singularity, as we now explain.
The frame variables $E_\alpha{}^i$
 satisfy equation (\ref{bsys}), which,
when specialized to the flat FL equilibrium point can be
solved to give
\[
	E_\alpha{}^i =\hat{E}_\alpha{}^i e^{\frac12(3\gam-2)t},
\]
where the $\hat{E}_\alpha{}^i$ are constants.
It follows that
\[
	\lim_{t\rightarrow-\infty} E_\alpha{}^i =0,
\]
assuming $\tfrac23<\gam<2$.
In addition, the constraints imply that the spatial Hubble gradient is 
zero ($r_\alpha=0$).
Thus in the Hubble-normalized $G_0$ state space,
the flat FL model is described by
an orbit that is past asymptotic to an equilibrium point on the silent
boundary.

$G_0$ cosmologies that have an isotropic singularity are described as 
follows.
Along a fixed timeline, the evolution is described by an orbit
that is past asymptotic to the flat FL equilibrium point on the silent
boundary.
The asymptotic time dependence for this class of models has been 
determined by systematically integrating the evolution equations 
(Lim \etal 2004).
The fact that the orbit is past asymptotic to the silent boundary means 
that the spatial dependence is asymptotically unrestricted, which 
manifests itself in the appearance of an arbitrary 3-metric in the
asymptotic expansion of the spacetime metric (see Lim \etal 2004).

\subsubsection*{Intermediate isotropization}

The flat FL equilibrium point F, given by (\ref{FL}), is a saddle point in 
the reduced Hubble-normalized state space for SH cosmologies. Thus, for 
any $\eps>0$ there is a family of orbits, corresponding to an open set of 
initial conditions, that pass through an $\eps$-neighbourhood of F but are 
not asymptotic to F.
For the corresponding cosmological models there will be a finite time 
interval during which the rate of expansion of the model is highly 
isotropic (i.e.~$\Sig \ll 1$)
and hence compatible with observational constraints.
This behaviour is called \emph{intermediate isotropization}, and occurs in 
models 
of all Bianchi types except Bianchi type I, in which case F is a sink.%
\footnote{Bianchi type I cosmologies undergo asymptotic isotropization as 
$t\rightarrow +\infty$.}

At this stage, intermediate isotropization for $G_0$
cosmologies has yet to be investigated in detail.

\section{The dynamics of nontilted SH cosmologies}\lb{sec_SH}

The largest class of cosmologies for which there is a comprehensive and 
quite detailed knowledge of the dynamics, is the class of nontilted SH 
cosmologies with $\Lambda\geq0$ and equation of state $p=(\gam-1)\rho$, 
with $\tfrac23<\gam<2$.
The reduced Hubble-normalized state space $\H_{\rm SH}$ is an unbounded 
subset of $\mathbb{R}^6$.
This class of cosmologies has two subsets referred to as class A and class 
B in the classification of Ellis \& MacCallum 1969.
We focus on the class A models, since they display a wider range of 
dynamical phenomena, while at the same time permitting a simpler choice of 
frame.
In describing the dynamics of this class of cosmologies, it is helpful to
make use of the hierarchy of invariant subsets of lower dimension that
arise from specializing the matter/energy content and the Bianchi type%
\footnote{The Bianchi types that occur in class A are I, II, VI$_0$,  
VII$_0$, VIII and IX, with the last two being of maximum generality.
We exclude Bianchi IX because models of this type can recollapse.}.
We list these invariant subsets in Table~\ref{tab_inv}.

\begin{table}[ht]
\caption{Dimension of invariant subsets}\lb{tab_inv}
\centering{
\begin{spacing}{1.1}
\begin{tabular}{lc}
\\
Invariant subset				& dimension
\\
\hline
zero cosmological constant $\Om_\Lambda=0$	&	5
\\
vacuum $\Om_m=0$				&	5
\\
Bianchi VI$_0$ and VII$_0$			&	5
\\
Bianchi II					&	4
\\
Bianchi I					&	3
\\
\hline
\end{tabular}\end{spacing}
}
\end{table}

We will discuss various dynamical phenomena, indicating how likely it is 
that each will occur.
We say that a dynamical behaviour is
\begin{itemize}
\item[i)]	\emph{generic}, if it occurs for all initial conditions 
		except for a set of measure zero%
\footnote{One can equivalently say ``for all orbits in the reduced 
Hubble-normalized state space except for a set of measure zero".},

\item[ii)]	\emph{typical}, if it occurs for a set of initial 
		conditions of positive measure, whose complement is also 
		of positive measure.

\item[iii)]	\emph{special}, it it occurs for a set of initial 
		conditions of measure zero.
\end{itemize}
One can say that generic behaviour has probability one, typical behavior 
has probability between zero and one, and special behaviour has 
probability zero.

We now list the various dynamical phenomena in Table~\ref{tab_phe} and 
discuss 
each in turn.

\begin{table}[ht]
\caption{Dynamical phenomena in nontilted SH cosmologies with $\Lambda>0$
and equation of state
$p=(\gam-1)\rho$, $\tfrac23 <\gam<2$.}\lb{tab_phe}
\centering{
\begin{spacing}{1.1}   
\begin{tabular}{cc}
\\
Dynamical behaviour	&	probability
\\
\hline
oscillatory singularity/Mixmaster dynamics	&	generic
\\
inflationary isotropization/future asymptotic to de Sitter & generic
\\
Weyl curvature domination			&	typical
\\
intermediate isotropization			& 	typical
\\
isotropic singularity				&	special
\\
\hline
\end{tabular}\end{spacing}
}
\end{table}

\subsection{Oscillatory singularity/Mixmaster dynamics}

Within $\H_{\rm SH}$, the Mixmaster dynamics is described by a past 
attractor, which is the union of the Kasner circle of equilibrium points 
and the vacuum Bianchi II orbits, first described by Ma \& Wainwright 
1992%
\footnote{This article is reprinted in Hobill \etal 1994. See also WE, 
Section 6.4.}.
The attractor contains infinite heteroclinic sequences, i.e.~infinite 
sequences of Kasner points joined by vacuum Bianchi II orbits.
A generic orbit that is asymptotic to the attractor shadows one of these 
heteroclinic sequences.
The evolution of the corresponding cosmological model towards the initial 
singularity is thus a non-terminating sequence of quasi-equilibrium Kasner 
states, punctuated by increasingly brief curvature transitions.
The early work on the attractor was based solely on heuristic local 
stability arguments and numerical simulations.
Recently, however, Ringstr\"{o}m 2001 proved%
\footnote{Ringstr\"{o}m's proof was given for the case of Bianchi IX 
cosmologies. A technical difficulty remains in extending the proof to 
Bianchi VIII cosmologies.}
the existence of the past attractor, using the Hubble-normalized evolution 
equations.
This result may have wider significance in view of the conjectures about 
the local past attractor for $G_0$ cosmologies that have recently been 
made (Uggla \etal 2003 and Section~\ref{sec_vac_num}).

As indicated in Table~\ref{tab_phe}, Mixmaster dynamics are generic.
Examples of special behaviour are provided by models which are past 
asymptotic to a specific Kasner solution and by models with an isotropic 
singularity.

\subsection{Inflationary isotropization}

Within $\H_{\rm SH}$, inflationary isotropization is simply described by 
the fact that the de Sitter equilibrium point (\ref{dS}), which describes
the de Sitter solution, is a global sink%
\footnote{This statement is essentially equivalent to the theorem of Wald 
1983, discussed in Section~\ref{sec_iso}.},
and hence forms the future attractor in the state space.
This behaviour is generic, and
special behaviour is restricted to the invariant set $\Om_\Lambda=0$.
The dimensionless scalars satisfy
\[
	\lim_{t\rightarrow+\infty} (\Sig,\ \Weyl,\ \Om_m)=0,\quad
	\lim_{t\rightarrow+\infty} \Om_\Lambda=1.
\]
The asymptotic time dependence can be obtained from
Lim \etal 2004 and Rendall 2003.

\subsection{Weyl curvature domination}

Generic orbits in the invariant subset $\Om_\Lambda=0$ escape to infinity, 
and satisfy
\be
\lb{Weyl_infty}
	\lim_{t\rightarrow+\infty} \Weyl = \infty.
\ee
This result holds for Bianchi VIII orbits if the equation of state 
satisfies $1\leq \gam<2$ (Horwood \& Wainwright 2004) and for Bianchi 
VII$_0$ orbits of $1<\gam<2$ (Wainwright \etal 1999).
In other words the Weyl curvature is dynamically dominant at late times.
From a mathematical perspective this result means that the invariant 
subset $\Om_\Lambda=0$ does not admit a future attractor.
Despite this fact, one can still determine the asymptotic dependence of 
the state vector as $t\rightarrow +\infty$
(Horwood \& Wainwright 2004).

The asymptotic behaviour in the invariant set  $\Om_\Lambda=0$ as 
$t\rightarrow +\infty$, as epitomized by (\ref{Weyl_infty}), has a 
significant effect on the dynamics in the full state space  
$\Om_\Lambda>0$, in that \emph{it approximates the intermediate dynamics} 
of a typical class of models with  $\Om_\Lambda>0$.
One can think of the asymptotic state of models with  $\Om_\Lambda=0$ 
acting as a \emph{generalized saddle} in the full state space, temporarily 
attracting orbits but eventually repelling them.
In summary, for a typical class of models with  $\Om_\Lambda>0$ there will 
be a finite epoch during which Weyl dominance occurs (i.e.~$\Weyl \gg 1$).

\subsection{Intermediate isotropization with $\Om_m\approx1$}

A cosmological model undergoes intermediate isotropization with 
$\Om_m \approx 1$ if there is a finite epoch during which the model is 
close to the flat FL model.
This behaviour is created by the local stability properties of the flat FL 
model.
We have seen that
the flat FL model has a multiple representation, as the flat FL 
equilibrium point F and as an 
unbounded orbit (see Section~\ref{sec_close_to__FL}, in particular, 
Figure~\ref{fig3.3}).
Orbits that are future asymptotic to F (i.e.~orbits that form the two 
dimensional stable manifold of F) satisfy
\be
\lb{lim_FL}
	\lim_{t\rightarrow+\infty}\Om_m=1,\quad
	\lim_{t\rightarrow+\infty}\Sig=0,\quad
	\lim_{t\rightarrow+\infty}\Weyl=0.
\ee
On the other hand orbits that are future asymptotic to the unbounded orbit 
satisfy%
\footnote{These orbits form the four dimensional Bianchi VII$_0$ invariant 
set with $\Om_\Lambda=0$.}
\be
\lb{lim_FL_infty}
	\lim_{t\rightarrow+\infty}\Om_m=1,\quad
        \lim_{t\rightarrow+\infty}\Sig=0,\quad
        \lim_{t\rightarrow+\infty}\Weyl=
	\begin{cases}
	L & \text{if $\gam=1$}\\
	+\infty & \text{if $1<\gam\leq \tfrac43$.}
	\end{cases}
\ee
We say that models that satisfy (\ref{lim_FL}) are 
\emph{asymptotic to flat FL in the strong sense}, 
while those satisfy (\ref{lim_FL_infty}) are 
\emph{asymptotic to flat FL in the weak sense}.
Both of these asymptotic states can act as (generalized) saddles in the 
full state space, leading to intermediate isotropization with 
$\Om_m\approx1$.
Thus
intermediate isotropization with $\Om_m\approx1$ in the full state space 
is 
a result of the occurrence of asymptotic isotropization with
$\lim\limits_{t\rightarrow+\infty}\Om_m=1$ in lower dimensional invariant
sets.

\begin{figure}[ht]
  \begin{center}  
    \epsfig{file=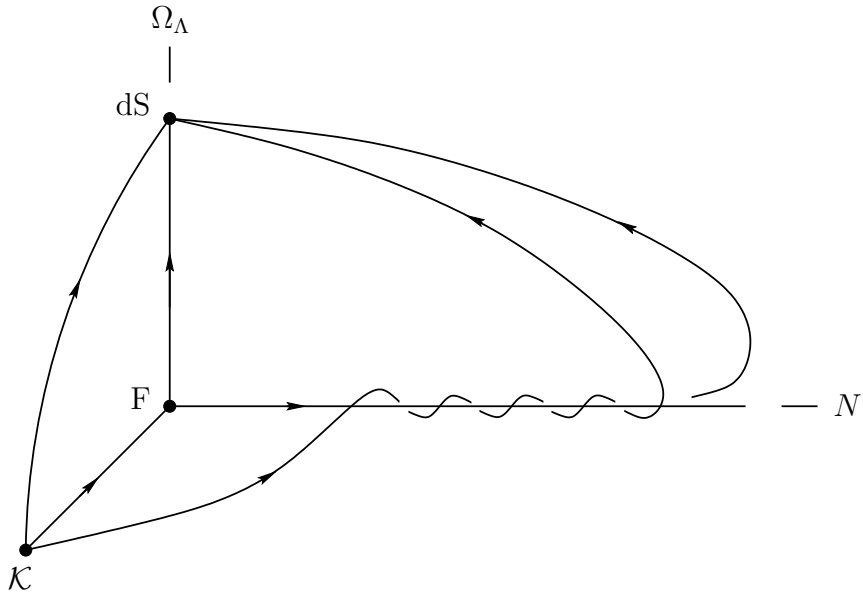,width=4.2in} 
\setlength{\unitlength}{1mm}
\begin{picture}(120,0)(0,0)
\put(27,75){\makebox(0,0)[b]{$\Om_\Lambda$}}
\put(23,25){\makebox(0,0)[b]{F}}
\put(7,1){\makebox(0,0)[b]{$\mathcal{K}$}}
\put(22,64){\makebox(0,0)[b]{dS}}
\put(117,24){\makebox(0,0)[b]{$N$}}
\end{picture}
\end{center}
\caption{Intermediate isotropization in the weak sense.}\lb{fig_saddle2}
\end{figure}

Intermediate isotropization in the weak sense with $\Om_m\approx1$ is
illustrated schematically in Figure~\ref{fig_saddle2}.
An orbit that is past asymptotic to a Kasner equilibrium point 
$\mathcal{K}$ spirals around the flat FL orbit, creating an epoch of 
intermediate isotropization with $\Om_m\approx1$, before approaching the 
de Sitter equilibrium point.

\subsection{Isotropic singularity}

Within $\H_{\rm SH}$, a cosmological model with an isotropic singularity 
is simply described by an orbit that is past asymptotic to the flat FL 
equilibrium point (\ref{FL}), i.e.~an orbit that lies in the unstable 
manifold of this equilibrium point. Recall that this manifold is of 
dimension 4 (see Table~\ref{tab_dim1}).
It follows that
\[
	\lim_{t\rightarrow -\infty} (\Sig,\ \Weyl,\ 
	\Om_\Lambda)=\mathbf{0},\quad
	\lim_{t\rightarrow -\infty} \Om_m=1
\]
for these models.
The asymptotic time dependence can be obtained from Lim \etal 2004.

\section{The vacuum past attractor and numerical 
simulations}\lb{sec_vac_num}

\subsection{The vacuum past attractor}

We now describe the attractor in the Hubble-normalized $G_0$ state space,
following 
the discussion in Uggla \etal 2003, but using a different choice of 
gauge.
One of the goals of that paper was to give a precise statement of the 
so-called BKL conjecture, by describing the past attractor in the 
Hubble-normalized state space.
\vskip4mm
\begin{minipage}[center]{0.8\linewidth}
\noindent
{\it The BKL conjecture.} For almost all cosmological solutions of 
Einstein's field equations, a spacelike initial singularity is 
\emph{vacuum-dominated, local and oscillatory}.
\end{minipage}
\vskip4mm

For cosmological models with a perfect fluid matter source, the phrase 
``vacuum-dominated," or, equivalently, ``matter is not dynamically 
significant," is taken to mean that the Hubble-normalized matter density 
$\Om_m$ tends to zero at the initial singularity.
One might then expect that the past attractor for cosmological models with 
a perfect fluid matter source would be the same as the past attractor for 
an idealized vacuum cosmological model.
It turns out (Uggla \etal 2003) that this expectation is unwarranted -- 
the 
peculiar velocity $v_\alpha$ plays a role in determining the past 
attractor. Nevertheless, partly in the interests of simplicity and also 
because the numerical simulations that have been recently performed 
(Garfinkle 2003) have been for vacuum models, we shall restrict our 
discussion here to the vacuum case.

In terms of Hubble-normalized variables and the separable volume gauge, 
the spacelike initial singularity in a $G_0$ cosmology is approached as 
$t\rightarrow -\infty$. We now define a \emph{silent initial singularity} 
to be one which satisfies
\begin{gather}
\lb{silent}
	\lim_{t\rightarrow -\infty} E_\alpha{}^i =0,\quad
	\lim_{t\rightarrow -\infty} r_\alpha=0,
\\
\lb{silent2}
	\lim_{t\rightarrow -\infty} \parb_\alpha \mathbf{Y}=0.
\end{gather}
More precisely, we require that (\ref{silent}) and (\ref{silent2}) are 
satisfied along typical timelines of $\mathbf{e}_0$.
One might initially think that the condition (\ref{silent2}) is a 
consequence of (\ref{silent}), since
\[
	\parb_\alpha \mathbf{Y} = E_\alpha{}^i 
	\frac{\ptl \mathbf{Y}}{\ptl x^i}.
\]
However the analysis of Gowdy solutions with so-called spikes%
\footnote{See Berger \& Moncrief 1993, Garfinkle \& Weaver 2003, Rendall 
\& Weaver 2001.}
shows that the partial derivatives $\ptl \mathbf{Y}/\ptl x^i$ can diverge 
as $t\rightarrow -\infty$.
Thus the requirement (\ref{silent2}) demands that $E_\alpha{}^i$ tend to 
zero sufficiently fast.
The asymptotic analysis of the Gowdy solutions shows that (\ref{silent2}) 
is satisfied along typical timelines even when spikes occur
(see Uggla \etal 2003, Section IV.A).
In more general $G_2$ cosmologies, however, the validity of 
(\ref{silent2}) is still open and further work is needed.

We now refer to Uggla \etal 2003, pages 9--10, for further evidence to 
justify making the following conjecture.
\vskip4mm
\begin{minipage}[center]{0.8\linewidth}
{\bf Conjecture 1}: For almost all cosmological solutions of Einstein's 
field equations, a spacelike initial singularity is silent.
\end{minipage}
\vskip4mm

Proving this conjecture entails establishing the limits 
(\ref{silent})--(\ref{silent2}).

\vspace{2mm}
We think of the evolution of the Hubble-normalized state vector 
$\mathbf{X}(t,x^i)$ \emph{for fixed $x^i$}, as being described by an orbit 
in a finite-dimensional Hubble-normalized state space.
As $t\rightarrow -\infty$, this orbit will be asymptotic to a local past 
attractor $\mathcal{A}^-$, which in accordance with the definition of 
silent initial singularity, will be contained in the silent boundary, 
defined by (\ref{silent_cond}).

The past attractor is based on the Kasner solutions.
We have seen that in the reduced Hubble-normalized state space for SH 
models, 
the Kasner solutions are represented by a sphere of equilibrium points.
In the full $G_0$ state space, these solutions are represented by orbits 
that are past asymptotic to Kasner equilibrium points on the silent 
boundary.
Thus we focus our attention on the
\emph{Kasner sphere $\mathcal{K}$ on the silent boundary},
defined by
\[
	A_\alpha=0,\quad N_{\alpha\beta}=0,\quad \Sig^2=1,
\]
in addition to (\ref{silent_cond}).
Each non-exceptional%
\footnote{The exceptional Kasner points describe the flat Kasner solution 
(the Taub form of Minkowski spacetime) and are characterized by the 
restriction
$\Sig_{\la\alpha}{}^\gam \Sig_{\beta\ra\gam} - \Sig_{\alpha\beta} =0$,
where $\la\ \ra$ denotes tracefree symmetrization.}
 equilibrium point on the Kasner sphere has one 
negative eigenvalue and hence has a one-dimensional unstable manifold into 
the past, which represents a SH vacuum solution of Bianchi type II on the 
silent boundary.
We refer to these orbits as \emph{curvature transitions}, since on each 
such orbit one degree of freedom of the spatial curvature is activated.%
\footnote{If one performs a spatial rotation to a shear eigenframe, one 
sees immediately
from the evolution equations for $N_{\alpha\beta}$
 that one diagonal component of $N_{\alpha\beta}$ is 
unstable into the past.}
These curvature transitions satisfy
\[
	\det(N_{\alpha\beta})=0,\quad
\Delta_N:= N_{\alpha\beta} N^{\alpha\beta} - (N_\alpha{}^\alpha)^2=0,\quad
	N_\alpha{}^\alpha \neq 0,
\]
corresponding to the fact that two eigenvalues of $N_{\alpha\beta}$ are 
zero.
In addition
\[	
	A_\alpha=0,
\]
and the Gauss constraint simplifies to
\[
	\Sig^2 + \tfrac{1}{12}(N_\alpha{}^\alpha)^2=1.
\]

We now make our second conjecture, motivated by our knowledge of SH 
dynamics.
\vskip4mm
\begin{minipage}[center]{0.8\linewidth}
{\bf Conjecture 2}: The local past attractor $\mathcal{A}^-$ for vacuum 
$G_0$ cosmologies with a silent initial singularity is
\[
	\mathcal{A}^- = \mathcal{K} \cup \mathcal{T}_N,
\]
where $\mathcal{K}$ is the Kasner sphere and $\mathcal{T}_N$ is the set of 
all curvature transitions in the silent boundary.
\end{minipage}
\vskip4mm

Establishing this conjecture entails proving the following limits:
\begin{gather}
\lb{A}
	\lim_{t\rightarrow-\infty} A_\alpha =0,
\\
	\lim_{t\rightarrow-\infty} \det(N_{\alpha\beta}) =0,
\\
\lb{N2}
	\lim_{t\rightarrow-\infty} [
	N_{\alpha\beta} N^{\alpha\beta} - (N_\alpha{}^\alpha)^2 ]=0.
\end{gather}
The fact that the sequence of transitions is non-terminating implies that
\[	\lim_{t\rightarrow-\infty} \Sig^2 \quad\text{and}\quad
	\lim_{t\rightarrow-\infty} N_{\alpha}{}^\alpha
\]
do not exist.
However it follows from the Gauss constraint (\ref{CGauss}) and equations
(\ref{silent}), (\ref{silent2}) and (\ref{N2}) that
\[
	\lim_{t\rightarrow-\infty} [
		\Sig^2 + \tfrac{1}{12} (N_{\alpha}{}^\alpha)^2]=1.
\]
In addition it follows from one of the other constraints in (\ref{esys}), 
the Codacci constraint%
\footnote{See (2.28) in Lim \etal 2004.},
and the limits (\ref{silent}), (\ref{silent2}) and (\ref{N2}) that
\be
\lb{eigen}
	\lim_{t\rightarrow-\infty} \eps^{\alpha\beta\gam}
		N_{\beta\d} \Sig_\gam{}^\d =0.
\ee
We thus expect that in a shear eigenframe, $N_{\alpha\beta}$ will be 
``asymptotically diagonal" as $t\rightarrow-\infty$.

The geometrical interpretation of the asymptotic evolution is as follows:
the orbit describing the evolution of the state vector 
$\mathbf{X}(t,x^i)$ for fixed $x^i$ approaches the Kasner sphere 
$\mathcal{K}$ in the silent boundary and then shadows
increasingly closely an infinite sequence 
of curvature transitions joining Kasner equilibrium points.

\subsection{Numerical simulations}

Some numerical simulations of the past asymptotic behaviour of vacuum 
$G_0$ cosmologies using the Hubble-normalized evolution equations 
(\ref{bsys})--(\ref{esys}) have been recently performed by Garfinkle 2003.
He found it convenient to use the deceleration parameter as a dynamical 
variable, with the defining equation (\ref{q}) forming an additional 
constraint. The evolution equation for $q$ has the form of a quasilinear 
diffusion equation, and it is the only evolution equation to contain 
second order spatial derivatives.

The Hubble-normalized evolution equations were solved numerically on a 
3-torus, i.e.~with periodic 
boundary conditions, using a Crank-Nicholson scheme.
The initial conditions were chosen to be of the form
\begin{gather*}
	E_\alpha{}^i= \psi^{-2} \d_\alpha{}^i,\quad
	r_\alpha = 0,\quad
	N_{\alpha\beta}=0,\quad
	A_\alpha = -2 \psi^{-3} \d_\alpha{}^i \ptl_i \psi,
\\
	\Sig_{\alpha\beta} = 
	\psi^{-6} 
	\text{diag}(\bar{\Sig}_{11},\bar{\Sig}_{22},\bar{\Sig}_{33})
	=\psi^{-6} \text{diag}
	\left(\begin{matrix}
	\ \  a_2 \cos y + a_3 \cos z +b_2 + b_3
\\
	\ \  a_1 \cos x - a_3 \cos z +b_1 - b_3
\\
	-a_1\cos x - a_2 \cos y -b_1 - b_2
	\end{matrix}\right),
\end{gather*}
where $a_\alpha$, $b_\alpha$ are arbitrary constants.

The Gauss constraint gives a nonlinear elliptic PDE for $\psi(x^i)$:
\[
	\nabla^2 \psi = \tfrac{3}{4} \psi^5 - \tfrac{1}{8}
\psi^{-7} \left[ \bar{\Sigma}_{11}{}^2 + \bar{\Sigma}_{22}{}^2 +
        \bar{\Sigma}_{33}{}^2
        \right]\ ,
\]
where $\nabla^2$ is the Euclidean Laplacian in $x$, $y$ and $z$. This PDE
has to be solved numerically
on a 3-torus
 to obtain explicit initial conditions.
The remaining constraints are satisfied identically.
Note that the above choice of $E_\alpha{}^i$ corresponds to a conformally 
flat initial 3-metric.

The numerical simulations show a brief transient epoch after which a 
sequence of curvature transitions takes place, following the usual BKL 
transition law, which we now describe.
The Kasner exponents $p_\alpha$ are related to the shear variables 
$\Sig_{\alpha\beta}$ in a shear eigenframe as follows:
\[
	\Sig_{\alpha\beta} = \text{diag}(3p_1-1,3p_2-1,3p_3-1).
\]
The exponents satisfy
\[
	p_1+p_2+p_3=1,\quad p_1{}^2+p_2{}^2+p_3{}^2=1,
\]
and thus can be expressed in terms of a single parameter $u$ according to
\[
	p_1 = - \frac{u}{1+u+u^2}\,,\quad
	p_2 = \frac{1+u}{1+u+u^2}\,,\quad
	p_3 =\frac{u(1+u)}{1+u+u^2}\ .
\]
The transition law between Kasner states can then be written in the form%
\footnote{See, for example, WE page 236.}
\[
	u \rightarrow \begin{cases}
			u-1, &\text{if $u \geq 2$}
			\\
			\dfrac{1}{u-1}, &\text{if $1<u<2$.}
		\end{cases}
\]
The numerical simulations also suggest that asymptotically 
$\Sig_{\alpha\beta}$ and $N_{\alpha\beta}$ have a common eigenframe in 
accordance with (\ref{eigen}).

The figures show the results of a numerical simulation
with $t=0$ initially, and then becoming negative, i.e. evolution towards 
the initial singularity ($t \rightarrow -\infty$).
Figures~\ref{fig_u1}a~and~\ref{fig_u2}a show the transitions of the 
Kasner parameter $u$ along two different timelines and 
Figures~\ref{fig_u1}b~and~\ref{fig_u2}b show the diagonal values of 
$N_{\alpha\beta}$ in the common asymptotic eigenframe, along the same 
timelines.
Overall, the simulations provide support for the limits 
(\ref{silent}) and (\ref{silent2}), and (\ref{A})--(\ref{eigen}) along 
typical timelines.

\begin{figure}
  \begin{center}
    \epsfig{file=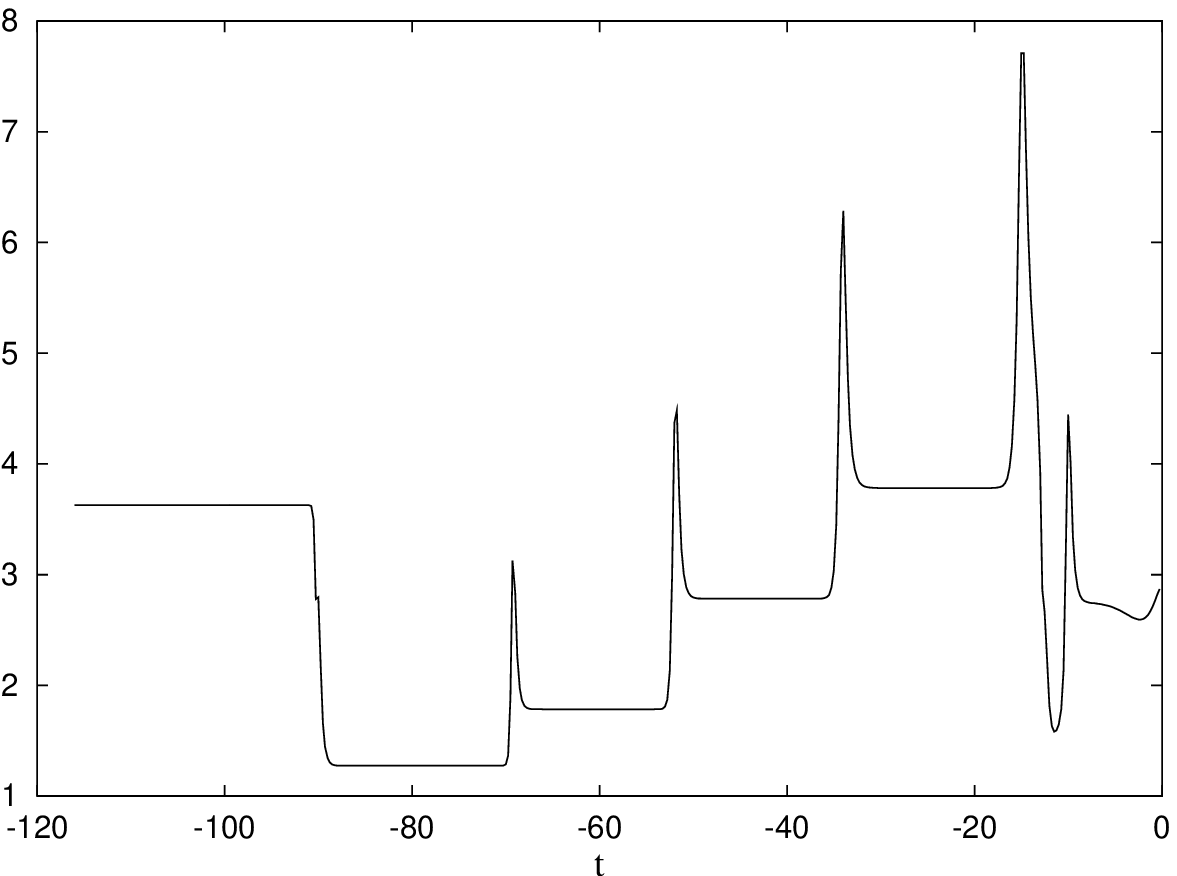}
    \epsfig{file=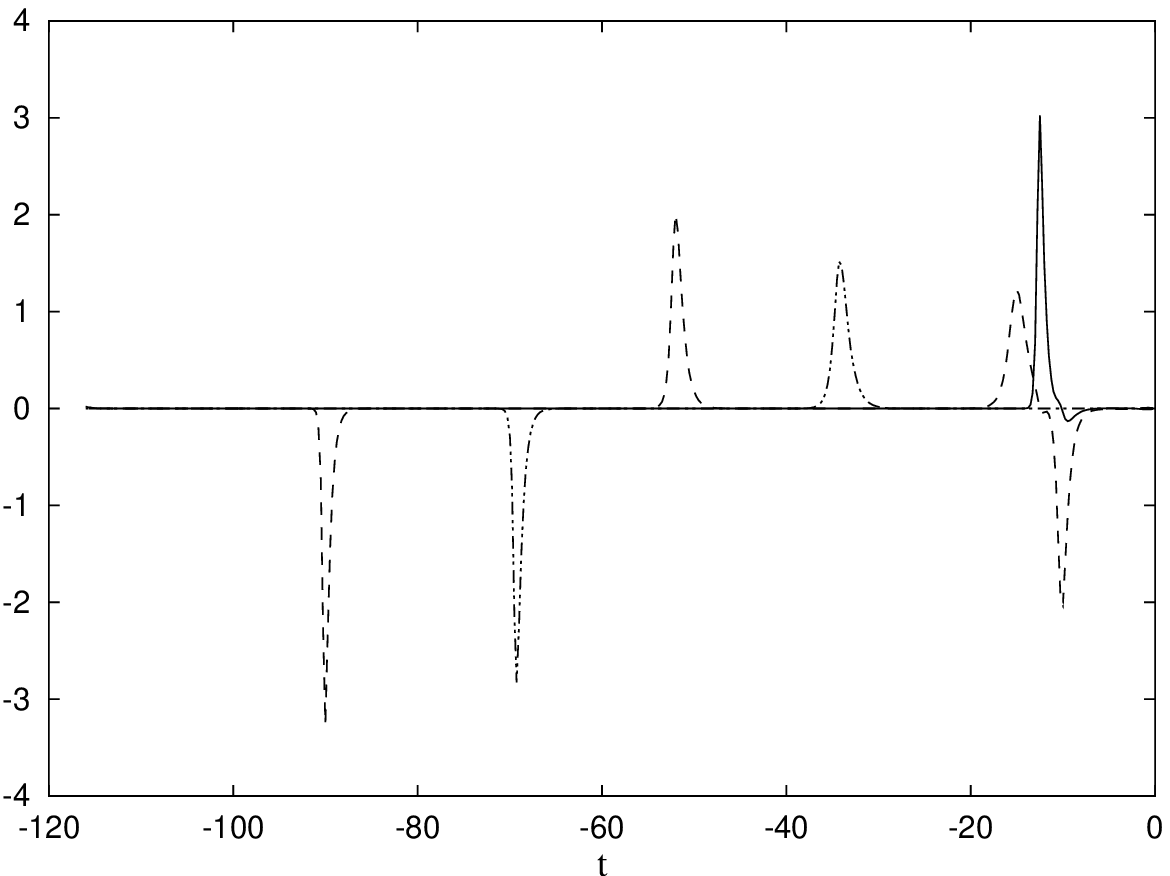}  
\end{center}
\caption{Transitions of the Kasner parameter $u$ and the 
diagonal values of $N_{\alpha\beta}$ in the common asymptotic eigenframe,
as $t\rightarrow -\infty$ along a fixed timeline, i.e.~for
fixed spatial coordinates $x^i$.}\lb{fig_u1}
\end{figure}

\begin{figure}
  \begin{center}
    \epsfig{file=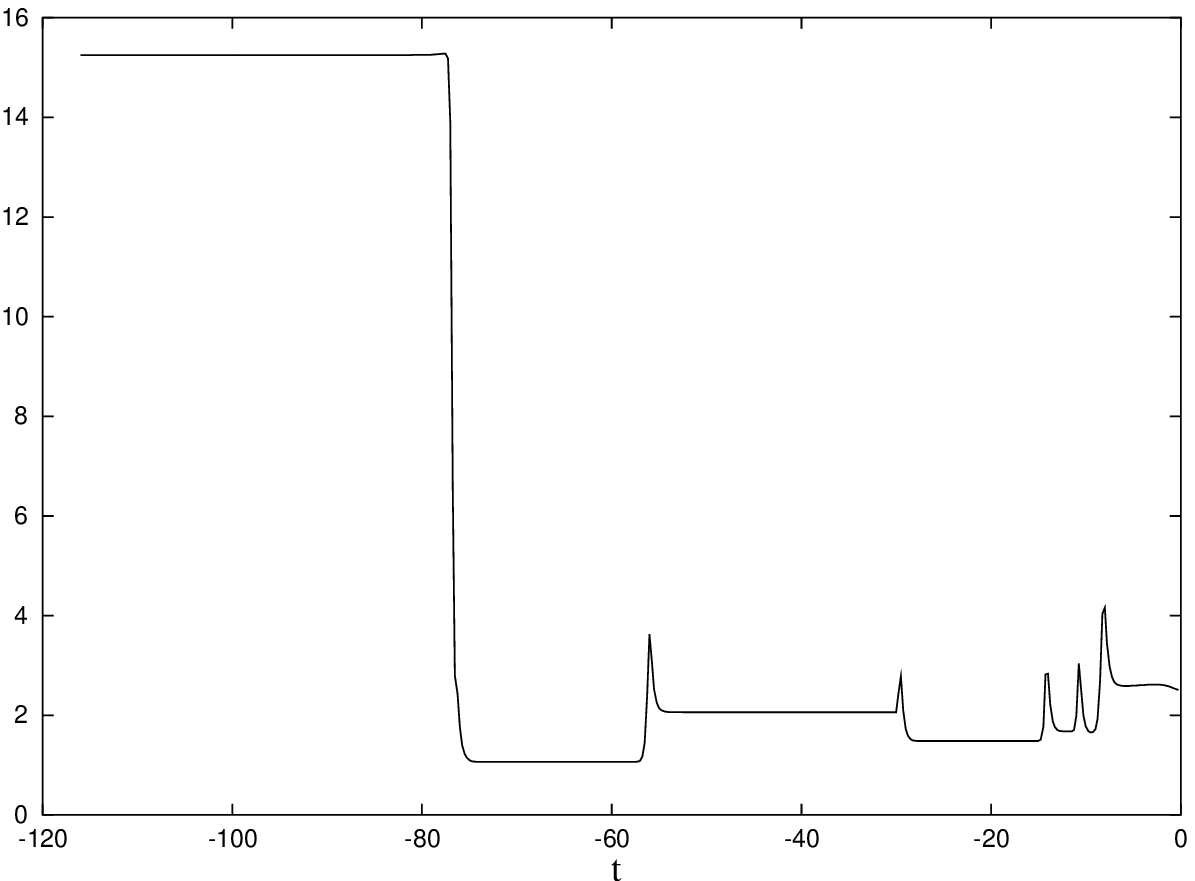}
    \epsfig{file=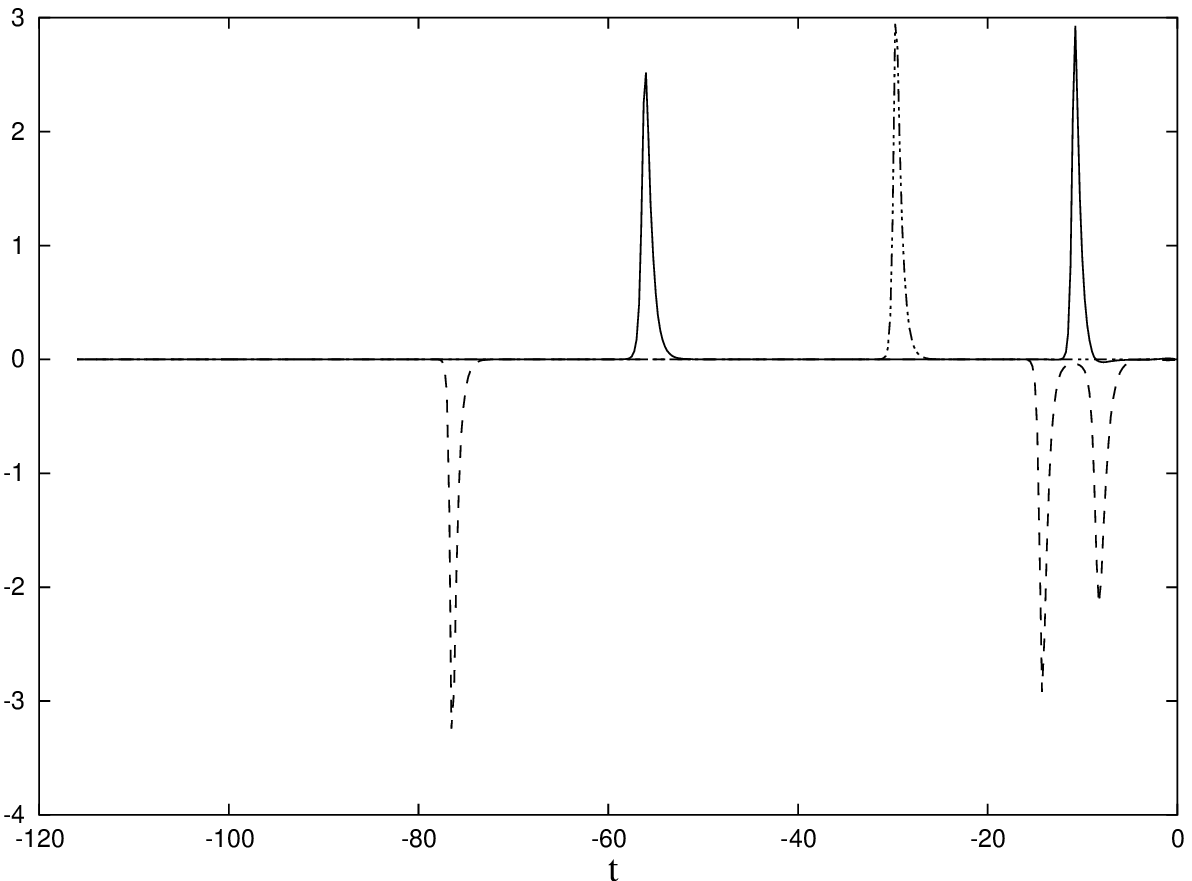} 
\end{center}
\caption{Transitions along a different fixed timeline.}\lb{fig_u2}
\end{figure}

The numerical simulation is incomplete in two ways.
First, the duration of the Kasner epochs in the simulation is far too 
short, due to 
the fact that the decay of the eigenvalues of $N_{\alpha\beta}$ to 
increasingly small values, which determines when the curvature transitions 
occur, is not described sufficiently accurately. The reason for this 
difficulty is that an arbitrarily chosen Fermi-propagated frame is not 
asymptotic to the eigenframe of $N_{\alpha\beta}$ as 
$t \rightarrow -\infty$.
Second, the simulation does not have high enough numerical resolution to 
correctly simulate spiky structures.
These structures are known to develop in $G_2$ cosmologies, on approach to 
the singularity (Berger \& Moncrief 1993), and we expect that similar 
structure will develop in $G_0$ cosmologies.
From our experience with numerical simulations of $G_2$ cosmologies, these 
spiky structures occur within the particle horizon of timelines where one 
of the eigenvalues of $N_{\alpha\beta}$ crosses zero
(Lim 2004).
As the particle 
horizon shrinks into the past, increasingly high numerical resolution is 
needed to simulate the spiky structures.
It may be prohibitively expensive to perform such simulations in $G_0$ 
cosmologies.

\section{A normalization for recollapsing models}\lb{sec_col}

The principal deficiency of Hubble-normalized variables is that they break 
down in an expanding cosmological model that reaches a maximum state of 
expansion ($H=0$), as is possible in an FL model with positive spatial 
curvature.
This difficulty can be addressed, at least for
FL models and SH models of Bianchi type IX, 
by using a modified normalization factor first proposed 
by Uggla (WE Section 8.5.2).

In the interests of simplicity we introduce the modified normalization 
procedure for the class of FL models with positive spatial curvature.
For these models
the normalization factor is defined by
\[
	D = \sqrt{H^2 + \tfrac16 \,{}^3\!R}\ ,
\]
where ${}^3\!R$ is the  curvature scalar of the metric induced on 
the hypersurfaces $t=\mbox{constant}$.
Since ${}^3\!R>0$, this factor is positive throughout the evolution, 
including at the time of maximum expansion $H=0$.
Dimensionless variables are defined in the usual way, using 
$D$ instead of $H$:
\[
	\tilde{\Om}_\Lambda = \frac{\Lambda}{3D^2},\quad
	\tilde{\Om}_m = \frac{\rho}{3D^2},\quad
	\tilde{H} = \frac{H}{D}.
\]
It follows that
\[
	\tilde{\Om}_m + \tilde{\Om}_\Lambda = 1.
\]
Introducing a time variable $\tilde{\tau}$ according to
\[
	\frac{dt}{d\tilde{\tau}} = \frac{1}{D},
\]
one obtains the following evolution equations for $\tilde{H}$ and 
$\tilde{\Om}_\Lambda$:
\begin{align*}
	\frac{d\tilde{H}}{d\tilde{\tau}} 
	&= -\tilde{q}(1-\tilde{H}^2),
\\
	\frac{d\tilde{\Om}_\Lambda}{d\tilde{\tau}}
	&= 2(1+\tilde{q})\tilde{H}\tilde{\Om}_\Lambda,
\end{align*}
where
\[
	\tilde{q} = \tfrac12(3\gam-2 - 3\gam \tilde{\Om}_\Lambda).
\]
We note that $\tilde{q}$ is related to the usual deceleration parameter 
$q$ according to
\[
	\tilde{q} = \tilde{H}^2 q,
\]
whenever $H \neq 0$.
The state space is bounded, being defined by the inequalities
\[
	-1 \leq \tilde{H} \leq 1,\quad
	0 \leq \tilde{\Om}_\Lambda \leq 1,
\]
and is shown in Figure~\ref{fig5.1}.
The sign of $\tilde{H}$ determines whether the model is expanding 
($\tilde{H}>0$) or collapsing ($\tilde{H}<0$).
The sign of $\tilde{q}$ determines whether the model is decelerating 
($\tilde{q}>0$) or accelerating.
The fixed points of the evolution equations are as follows:
\begin{alignat*}{3}
	\tilde{\Om}_\Lambda &=0,& \tilde{H}^2 &=1, &
	&\quad\text{the flat FL model ${\rm F}_\pm$,}
\\
	\tilde{\Om}_\Lambda &=1,& \tilde{H}^2 &=1, &
	&\quad\text{the de Sitter model ${\rm dS}_\pm$,}
\\
	\tilde{\Om}_\Lambda &=\frac{3\gam-2}{3\gam}, &\ \tilde{H}&=0, &
	&\quad\text{the Einstein static model E.}
\end{alignat*}
The points ${\rm F}_+$ and ${\rm dS}_+$ represent expanding models 
($\tilde{H}=1$) while ${\rm F}_-$ and ${\rm dS}_-$ represent the time 
reversed models ($\tilde{H}=-1$).
The orbits ${\rm F}_+ \rightarrow {\rm F}_-$ represent models that expand 
from a big-bang singularity and then recollapse to a future singularity.
The orbits ${\rm F}_+ \rightarrow {\rm dS}_+$ represent models that expand 
indefinitely from a big-bang singularity, enter an accelerating epoch, and 
approach de Sitter at late times.

\begin{figure}[ht]
  \begin{center}
    \epsfig{file=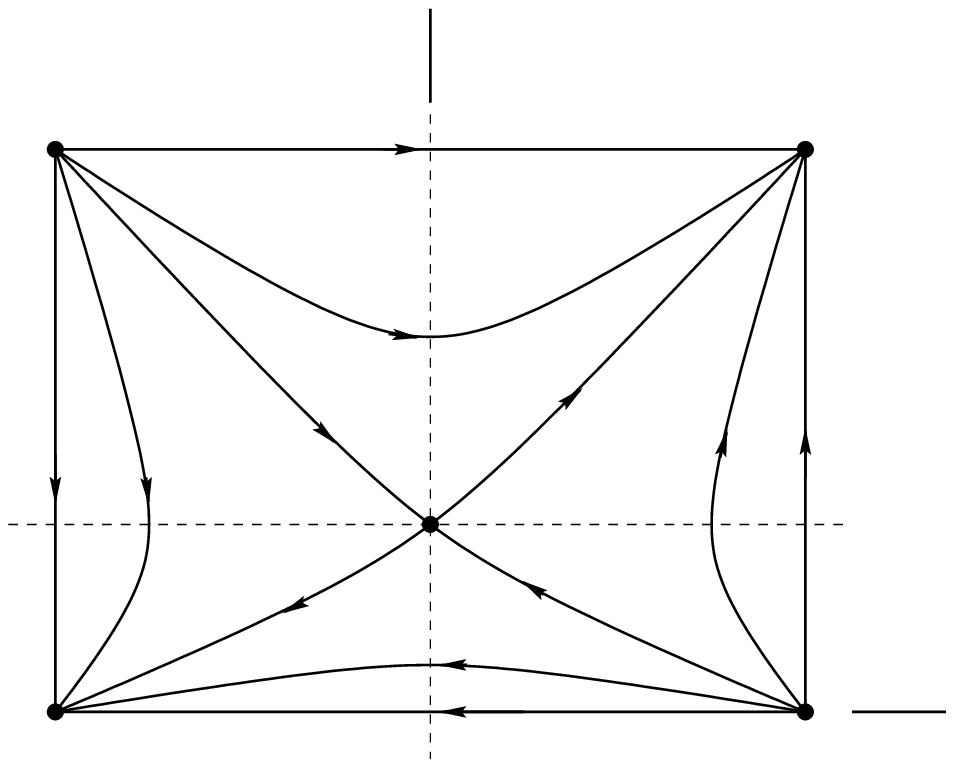,width=4.5in}
\setlength{\unitlength}{1mm}
\begin{picture}(120,0)(0,0)
\put(95,70){\makebox(0,0)[b]{${\rm dS}_+$}}
\put(17,70){\makebox(0,0)[b]{${\rm dS}_-$}}
\put(35,70){\makebox(0,0)[b]{contracting}}
\put(75,70){\makebox(0,0)[b]{expanding}}
\put(53,33){\makebox(0,0)[b]{E}}
\put(105,28){\makebox(0,0)[b]{$\tilde{q}=0$}}
\put(110,45){\makebox(0,0)[b]{accelerating}}
\put(110,20){\makebox(0,0)[b]{decelerating}}
\put(55,85){\makebox(0,0)[b]{$\tilde{\Om}_\Lambda$}}
\put(55,0){\makebox(0,0)[b]{$\tilde{H}=0$}}
\put(95,5){\makebox(0,0)[b]{${\rm F}_+$}}
\put(95,0){\makebox(0,0)[b]{(flat FL, expanding)}} 
\put(17,5){\makebox(0,0)[b]{${\rm F}_-$}}
\put(15,0){\makebox(0,0)[b]{(flat FL, contracting)}}   
\put(110,9){\makebox(0,0)[b]{$\tilde{H}$}}
\end{picture}
\end{center}
\caption{State space for the FL cosmologies with positive and zero spatial 
curvature,
$\Lambda\geq0$ 
and equation of state
$p=(\gam-1)\rho$, $\tfrac23<\gam<2$.}\lb{fig5.1}
\end{figure}

Figure~\ref{fig5.1} illustrates the central role played by the Einstein 
static model within the class of FL cosmologies with positive and zero 
spatial curvature.
The fixed point E is a saddle point whose stable and unstable manifolds 
are one-dimensional. 
The unstable manifold describes a one-parameter family of cosmologies that 
are past asymptotic to the Einstein static model, and hence do not have an 
initial singularity. 
These cosmological models have recently been proposed by Ellis \& Maartens 
2004
as a possible description of the early universe, a so-called emergent 
universe.

We note that this normalization procedure can be generalized to the class 
of nontilted SH cosmologies of Bianchi type IX. The normalization factor 
is (WE, equation (8.6))
\[
	D = \sqrt{H^2 + \tfrac14(n_{11} n_{22} n_{33})^{\frac23}}\ ,
\]
where $n_{\alpha\beta}=\text{diag}(n_{11},n_{22},n_{33})$.
The resulting state space, which gives a unified description of the 
dynamics of nontilted SH cosmologies of Bianchi type IX, VII$_0$, II and 
I, has not been explored in detail.
In addition, 
a variation of this normalization has been given by Heinzle \etal 2004, 
and has been used to give a detailed analysis of the nontilted SH 
cosmologies of Bianchi type IX that are locally rotationally symmetric.

\section{Conclusion}\lb{sec_conclusion}

The dynamical systems approach has two key attributes.
Firstly, it provides a geometric framework for analyzing the space of 
cosmological models. Secondly, it provides a system of first order 
evolution equations, whose specific form depends on the choice of gauge.
This system of equations is potentially useful for giving a rigorous 
analysis of the asymptotic dynamics of cosmological models, and also for 
performing numerical simulations.
The utility of this approach has been amply demonstrated as regards the SH 
cosmologies, and there are indications that it may be equally useful as 
regards the $G_0$ cosmologies.

\section*{Acknowledgements}

This work has been influenced significantly by our collaboration with 
Claes Uggla and Henk van Elst. It is a pleasure to acknowledge the many 
discussions that we have had over the past few years. We also thank George 
Ellis for his continued interest and encouragement, and David Garfinkle 
for providing Figures~\ref{fig_u1} and~\ref{fig_u2}, and for helpful 
discussions on the 
numerical simulations.
J.W. expresses his appreciation to Lars Andersson and Gregory Galloway and 
the Department of Mathematics at the University of Miami for their 
hospitality during the Miami Waves Conference in January 2004. This work 
was partially supported by a grant to J.W. from the Natural Sciences and 
Engineering Research Council of Canada.

\end{spacing}
\end{document}